\documentclass[aps,a4paper,amsmath,amssymb,twocolumn,
floatfix,superscriptaddress,showkeys]{revtex4-1}
\usepackage{bm,amsmath,amssymb,amsfonts}
\usepackage[dvips]{graphicx}
\usepackage[utf8]{inputenc}
\usepackage[normalem]{ulem}
\usepackage{color}
\usepackage{xcolor}
\usepackage{hyperref}

\hypersetup{
    pdfnewwindow=true,     
    colorlinks=true,       
    linkcolor=magenta,     
    citecolor=blue,        
    filecolor=blue,        
    urlcolor=blue          
}
\begin{document}
\title{Proximity-induced flat bands and topological properties 
in a decorated diamond chain}
\author{K Shivanand Thakur}
\author{Vihodi Theuno}
\affiliation{Department of Physics, School of Sciences, 
Nagaland University, Lumami 798627, Nagaland, India} 
\author{Amrita Mukherjee}
\affiliation{Department of Condensed Matter Physics and Materials Science, 
Tata Institute of Fundamental Research, Mumbai 400005, India} 
\author{Biplab Pal}
\thanks{Corresponding author}
\email[E-mail: ]{biplab@nagalanduniversity.ac.in}
\affiliation{Department of Physics, School of Sciences, 
Nagaland University, Lumami 798627, Nagaland, India}
\date{\today}
\begin{abstract}
In the present study, we propose a unique scheme to generate and control multiple 
flat bands in a decorated diamond chain by using a strain-induced proximity effect 
between the diagonal sites of each diamond plaquette. This is in complete contrast 
to the conventional diamond chain, in which the interplay between the lattice topology 
and an external magnetic flux leads to an extreme localization of the single-particle 
states, producing the flat bands in the energy spectrum. Such a strain-induced proximity 
effect will enable us to systematically control one of the diagonal hoppings in the 
decorated diamond chain, which will lead to the formation of both gapless and gapped 
flat bands in the energy spectrum. These gapless or gapped flat bands have been 
corroborated by the computation of the compact localized states amplitude distribution 
as well as the density of states of the system using a real space calculation. We have 
also shown that these flat bands are robust against the introduction of small 
amounts of random onsite disorder in the system. In addition to this, we have also 
classified the nontrivial topological properties of the system by calculating the 
winding numbers and edge states for the gapped energy spectrum. These findings could be 
easily realized experimentally using the laser-induced photonic lattice platforms. 
\end{abstract}
\maketitle
\section{Introduction}
\label{sec:Intro}
Understanding the nature of exotic electronic states plays a pivotal role in building 
modern quantum technology-based devices. In quantum systems, the characteristics of the 
electronic states govern the quantum transport properties, which is a crucial component of 
quantum information processing and quantum communication. In general, there exist two 
different kinds of electronic states in quantum systems, viz., \emph{extended} 
states and \emph{localized} states. The first one is generally attributed to any wave-like 
phenomena in a periodic medium and known as the Bloch states~\cite{Bloch-1929}, whereas the 
latter one describes the absence of diffusion of waves in a completely disordered or chaotic 
medium and manifests in the phenomena of Anderson localization~\cite{Anderson-prb-1958,
Ramakrishnan-prl-1979}. Later on, it was found that we can have exceptions to the above 
scenarios, where one can also have highly localized states even in a perfectly periodic 
system in the absence of any disorder. Such disorder-free localized states appearing in 
periodic lattice models are known as the flat-band (FB) states~\cite{FB-review-Flach-2018,
FB-review-Vicencio-2021}. 

There are various reasons due to which these FB states emerge in different lattice 
models, such as the local topology of the lattice model~\cite{Sutherland-prb-1986}, 
orbital frustration~\cite{Trivedi-prb-2022}, influence of external magnetic 
flux~\cite{Dias-prb-2011,Pal-prb-2018,Pal-jpcm-2025}, influence of electric 
field~\cite{Flach-prb-2018,Flach-prr-2021}, hopping frustration~\cite{Nat-Phys-2024}, 
and so on. At the FB energies, the kinetic energy of the particle is quenched, making it 
completely immobile, and hence it becomes highly localized. Due to this, the corresponding 
electronic wavefunction is caged over a couple of unit cells of the lattice geometry, and 
outside this region, it abruptly decays to zero. One can exactly construct the distribution 
of the wavefunction amplitudes for such localized states at the FB energies, which are known as 
the compact localized states (CLS)~\cite{Ajith-prb-2017,Singular-FB-Rhim-2021,Pal-prb-2018,Pal-jpcm-2025}. 
\begin{figure}[ht]
\centering
\includegraphics[clip, width=\columnwidth]{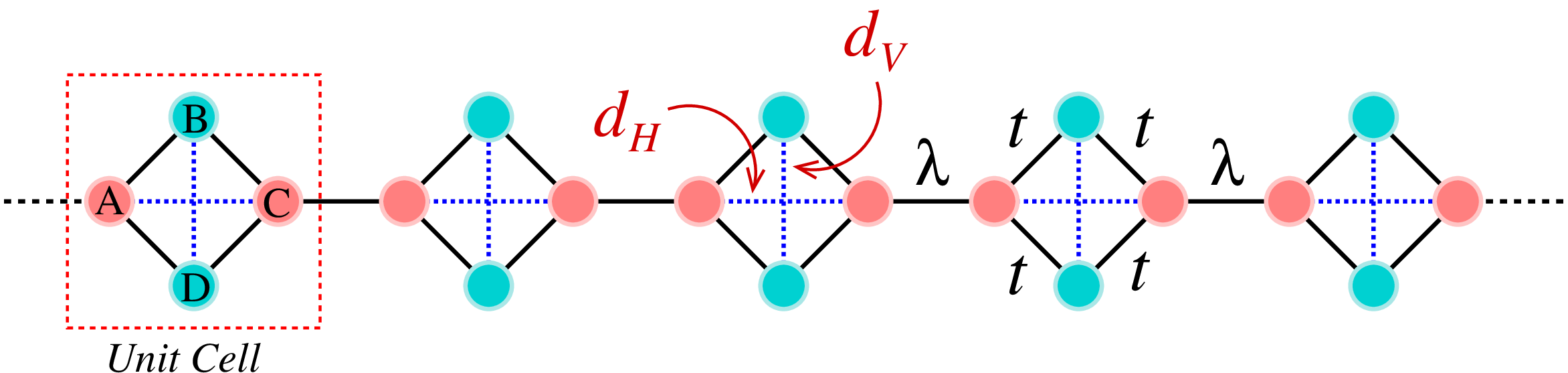}
\caption{Schematic diagram of a quasi-one-dimensional decorated 
diamond lattice. The segment inside the dotted box indicates 
the unit cell of the lattice structure, consisting of four atomic 
sites. $t$ denotes the coupling along the periphery of each 
diamond (rhombic) structure, while $d_H$ and $d_V$ depict 
the diagonal couplings along the \emph{horizontal} and 
\emph{vertical} directions, respectively, inside a diamond 
structure. $\lambda$ represents the coupling in between two 
consecutive diamond structures.} 
\label{fig:Lattice-model}
\end{figure}

Such FB states in the band structure of lattice models have attracted a lot of interest 
in recent times because they provide an excellent platform to explore various intriguing 
strongly correlated phenomena like the quantum Hall effect in the absence of a magnetic 
field~\cite{Wen-prl-2011,Das-Sarma-prl-2011,Neupert-prl-2011}, FB superconductivity~\cite{FB-superconductivity-1,
FB-superconductivity-2,FB-superconductivity-3,FB-superconductivity-4}, inverse Anderson localization 
transition~\cite{Inverse-Anderson-transition-1,Inverse-Anderson-transition-2,Inverse-Anderson-transition-3}, 
to name a few among others. These theoretical investigations on FB physics have been further 
accelerated by their experimental realization in the field of photonics~\cite{Rodrigo-prl-2015,
Sebabrata-prl-2015,Sebabrata-prl-2018,Rodrigo-prl-2022}. The rapid technological advancement in 
the fabrication and characterization techniques used in the field of photonics has empowered us 
to fabricate tailor-made photonic lattice structures with arbitrary lattice geometries and 
experimentally observe the localized FB states in those photonic lattice settings. Such techniques 
have enabled us to go beyond the regular lattice geometries and to explore FB states 
featuring in more complicated unusual lattice structures like super-honeycomb lattice~\cite{Chen-Suepr-Honeycomb-AOM-2020}, 
super-Kagome lattice~\cite{Chen-Super-Kagome-OL-2023}, decorated Lieb lattice~\cite{Denz-Extended-Lieb-FB,
Bhatta-Pal-prb-2019}, and even in fractal-like lattices~\cite{Pal-Saha-prb-2018,Hahafi-fractal-FB,
Chen-fractal-FB-1,Chen-fractal-FB-2,Biswas-fractal-FB}. Of late, an exciting addition to the 
study of FB physics has been the observation of FB localization and topological 
edge states in topolectrical circuits~\cite{Zhang-EC-FB-1,Zhang-EC-FB-2,Zhang-EC-FB-3,Flach-EC-FB}. 

The study of these FB states is not only important from the context of fundamental physics, 
but also it has various potential technological applications, such as quantum communication~\cite{FB-quantum-commun}, 
controlled optical transmission~\cite{Sebabrata-prl-2018,Rodrigo-prl-2022}, 
image processing and precise optical measurement~\cite{Sebabrata-prl-2015}, 
diffraction-free propagation of light~\cite{Rodrigo-prl-2015,Sebabrata-prl-2015}, 
telecommunication and sensing devices~\cite{Rodrigo-prl-2015}, and many more~\cite{Flach-Nanophoton-2024}. 
Lattice models are widely used in physics for a variety of reasons, including approximating 
actual physical systems and offering useful settings for interesting theoretical investigation. 
Thus, simple lattice models hosting tunable FBs in their band structure are always significant for 
both technological applications and the investigation of novel fundamental physics. 
In the present study, we have considered one such simple quasi-one-dimensional decorated diamond 
lattice (also called a rhombic lattice) model (see Fig.~\ref{fig:Lattice-model}) 
to explore the FB physics as well as the topological properties. 

The conventional diamond lattice structures have been explored earlier in the context of extreme 
localization using an external magnetic flux. This phenomenon is known as the Aharonov-Bohm (AB) 
caging effect~\cite{Vidal-prl-2000,Vidal-prl-1998}, where the particle is completely localized, 
giving rise to zero transport. In recent times, AB caging phenomena have been experimentally 
realized using photonic lattices~\cite{Sebabrata-prl-2018,Rodrigo-prl-2022} and also using a 
synthetic lattice in the momentum space of ultracold atoms with tailored gauge 
fields~\cite{Inverse-Anderson-transition-3}. Also, lately the phenomenon of AB caging has been 
theoretically explored for extended diamond chains~\cite{Amrita-jpcm-2021,Sougata-PhysicaB-2024} 
and even for fractal structures built out of diamond-shaped structures~\cite{Biplab-pssrrl-2025}. 
However, in all these studies, the extreme localization effect is induced by an external magnetic flux. 
In contrast to that, in the present study, we have examined a variant of the traditional diamond chain, 
where we have shown that FB localization can be tuned using a proximity effect. We have also demonstrated 
that the topological edge states in this lattice structure are caused by this proximity effect. 
A similar diamond-necklace chain, but without the diagonal couplings, has been studied previously in 
the context of compact localized boundary states~\cite{CM-Smith-Quantum-Front-2023} and Hilbert space 
fragmentation~\cite{Dias-prb-2023}. Also, in a very recent study, a similar antiferromagnetic 
diamond chain has been demonstrated to act as an efficient spin filter~\cite{Debjani-AoP-2025}.

Our results and analysis follow next. The article is structured as follows: In Sec.~\ref{sec:model}, 
we describe the construction and parameters of our lattice model and the tight-binding framework used 
to extract the results. The characterization of the proximity-induced FB states is demonstrated in 
Sec.~\ref{sec:Flat-bands}, which includes the construction of the CLS, computation of the 
density of states, and the effect of disorder on the FB states. In Sec.~\ref{sec:topo-properties},
we present the results related to the topological properties of the system, which includes calculation 
of the edge states and the winding number. Finally, we summarize the main findings of this study and 
its potential future scope in Sec.~\ref{sec:conclu}.
\section{Description of the model}
\label{sec:model} 
In this work, we have considered a quasi-one-dimensional decorated diamond lattice that grows along 
one direction only (see Fig.~\ref{fig:Lattice-model}). The building block of this lattice model is a 
diamond (rhombic)-shaped structure that consists of four atomic sites, namely, $A$, $B$, $C$, and $D$, 
as shown in Fig.~\ref{fig:Lattice-model}. The electronic properties of the single-electron states in this 
lattice model can be described by the tight-binding Hamiltonian written in the Wannier basis of the following form:
\begin{widetext}
\begin{align}
\bm{H} = \sum_{n}\Big[
\underbrace{ \sum_{i}\varepsilon_{i} 
c_{n,i}^{\dagger}c_{n,i}
}_\textit{Onsite energy}
+ 
\underbrace{ \sum_{i,j} \Big(\mathcal{T}_{i,j} 
c_{n,i}^{\dagger}c_{n,j} + \textrm{h.c.}\Big) 
}_\textit{Intra-cell hoppings}
+ 
\underbrace{ \sum_{\langle i,j \rangle}\Big(\lambda_{i,j} 
c_{n,i}^{\dagger}c_{n+1,j} + \textrm{h.c.}\Big) 
}_\textit{Inter-cell hoppings}
\Big],
\label{eq:hamilton-wannier}
\end{align}
\end{widetext} 
where $\varepsilon_{i}$ represents the onsite energy of an electron to be placed at the $i$-th site 
($i \in A,B,C,D$) in the $n$-th unit cell of the lattice. $\mathcal{T}_{i,j}$ represents the coupling 
(hopping amplitude) between the $i$-th and $j$-th sites within the same unit cell. $\mathcal{T}_{i,j}$ 
can take three possible values, viz., $t$ (when $i$ and $j$ are the two neighboring sites along 
the periphery of each diamond structure, \textit{i.e.}, in between a red and a cyan site), $d_H$ (when $i$ and $j$ 
represent the two neighboring sites along the \emph{horizontal} diagonal of each diamond structure, 
\textit{i.e.}, in between two red sites), and $d_V$ (when $i$ and $j$ represent the two neighboring sites 
along the \emph{vertical} diagonal of each diamond structure, \textit{i.e.}, in between two cyan sites) 
[see Fig.~\ref{fig:Lattice-model} for a better clarification]. $\lambda_{i,j} \equiv \lambda$ indicates the coupling 
between two neighboring sites $i$ and $j$ from the two neighboring unit cells, namely, $n$-th and $(n+1)$-th cells. 
$c_{n,i}^{\dagger}(c_{n,i})$ represents the creation (annihilation) operator for an electron at the $i$-th 
site in the $n$-th cell. We remark that, for our study, we have considered the spinless non-interacting electron 
formalism. 

To compute the electronic band structure of the system, we transform the Hamiltonian in Eq.~\eqref{eq:hamilton-wannier} 
from the Wannier basis description to the Bloch basis description using a discrete Fourier transform, which leads to 
\begin{equation}
\bm{H} = \sum_{k} \bm{\tilde{c}}^{\dagger}_{k} 
\bm{\mathcal{H}}(k)\bm{\tilde{c}}_{k},
\label{eq:hamilton-k-space} 
\end{equation}
where $\bm{\mathcal{H}}(k)$ matrix is given by,  
\begin{align}
\bm{\mathcal{H}}(k) =
\left[\def\arraystretch{1.8} \begin{matrix}
\varepsilon_{A} &  t  &  (d_{H} + \lambda e^{ika})  &  t \\
t  &  \varepsilon_{B}  &  t  &  d_{V} \\
(d_{H} + \lambda e^{-ika})  &  t  &  \varepsilon_{C}  &  t\\
t  &  d_{V}  &  t  &  \varepsilon_{D} \\
\end{matrix}\right]
\label{eq:hamilton-bloch}
\end{align}
and 
$\bm{\tilde{c}}^{\dagger}_{k}= \left(\begin{matrix}
c^{\dagger}_{k,A}  &  c^{\dagger}_{k,B}  
&  c^{\dagger}_{k,C} & c^{\dagger}_{k,D} 
\end{matrix}\right)$. 
For the perfectly ordered system, we set the onsite potentials at all the sites $\varepsilon_{i}=0$. 
However, later we also investigate the effect of the random onsite disorder on the electronic 
properties of this lattice model, and in that case, we choose $\varepsilon_{i}$ randomly within 
an interval$[-\frac{\Delta}{2},\frac{\Delta}{2}]$, where $\Delta$ controls the strength of the 
random disorder. Throughout our calculation, we set the hopping parameters $\lambda=t=1$. We play 
with the hopping parameters $d_H$ and $d_V$, which can be appropriately controlled by the 
so-called \emph{proximity effect}, to extract interesting FB features offered by this lattice model. 
This is elaborated in detail in Sec.~\ref{sec:Flat-bands}.  
\section{Characterization of the proximity-induced flat bands}
\label{sec:Flat-bands} 
In this section, we discuss how to engender and control multiple FBs in this lattice structure. 
FBs appear in the band structure of certain lattice models or in real materials mainly as a result 
of destructive interference of the electron hoppings. In this lattice model, we realize such a 
destructive interference effect by controlling certain hopping parameters in the system with a 
proximity effect. This gives us an effective approach to manipulate multiple FBs in this lattice 
structure, as discussed below. 
%
\begin{figure}[ht]
\centering
\includegraphics[clip, width=0.49\columnwidth]{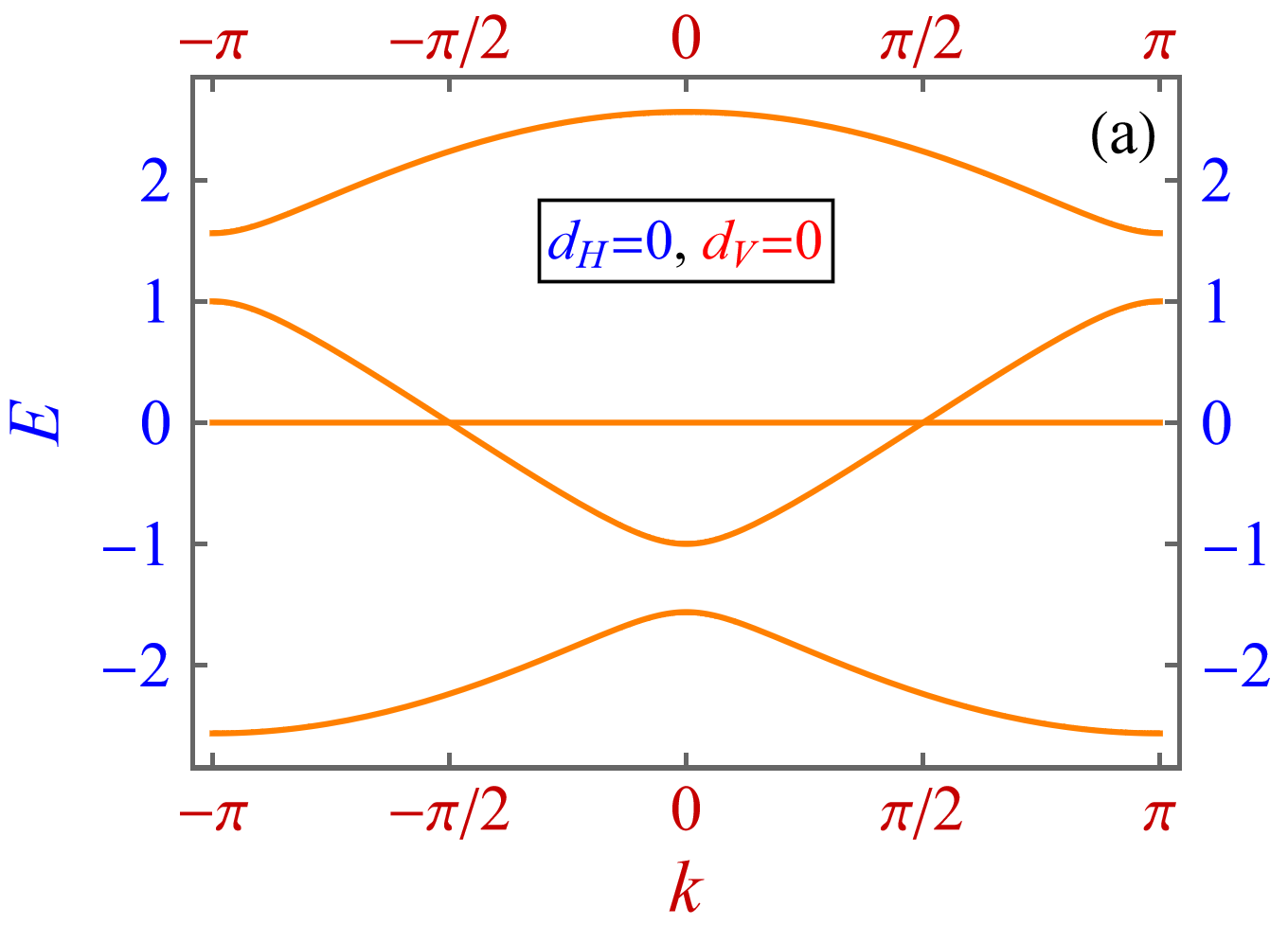}
\includegraphics[clip, width=0.49\columnwidth]{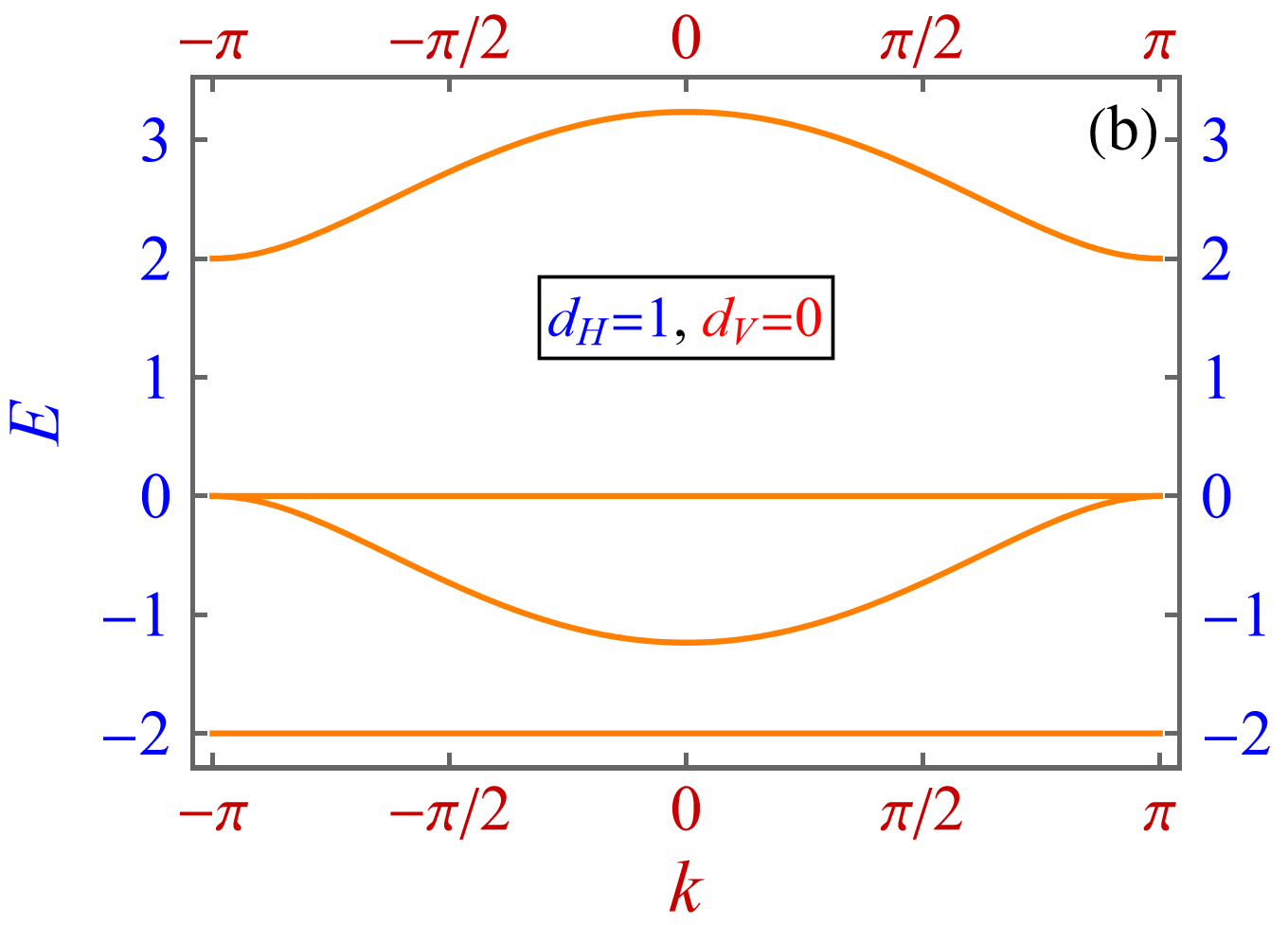}
\vskip 0.4cm
\includegraphics[clip, width=0.49\columnwidth]{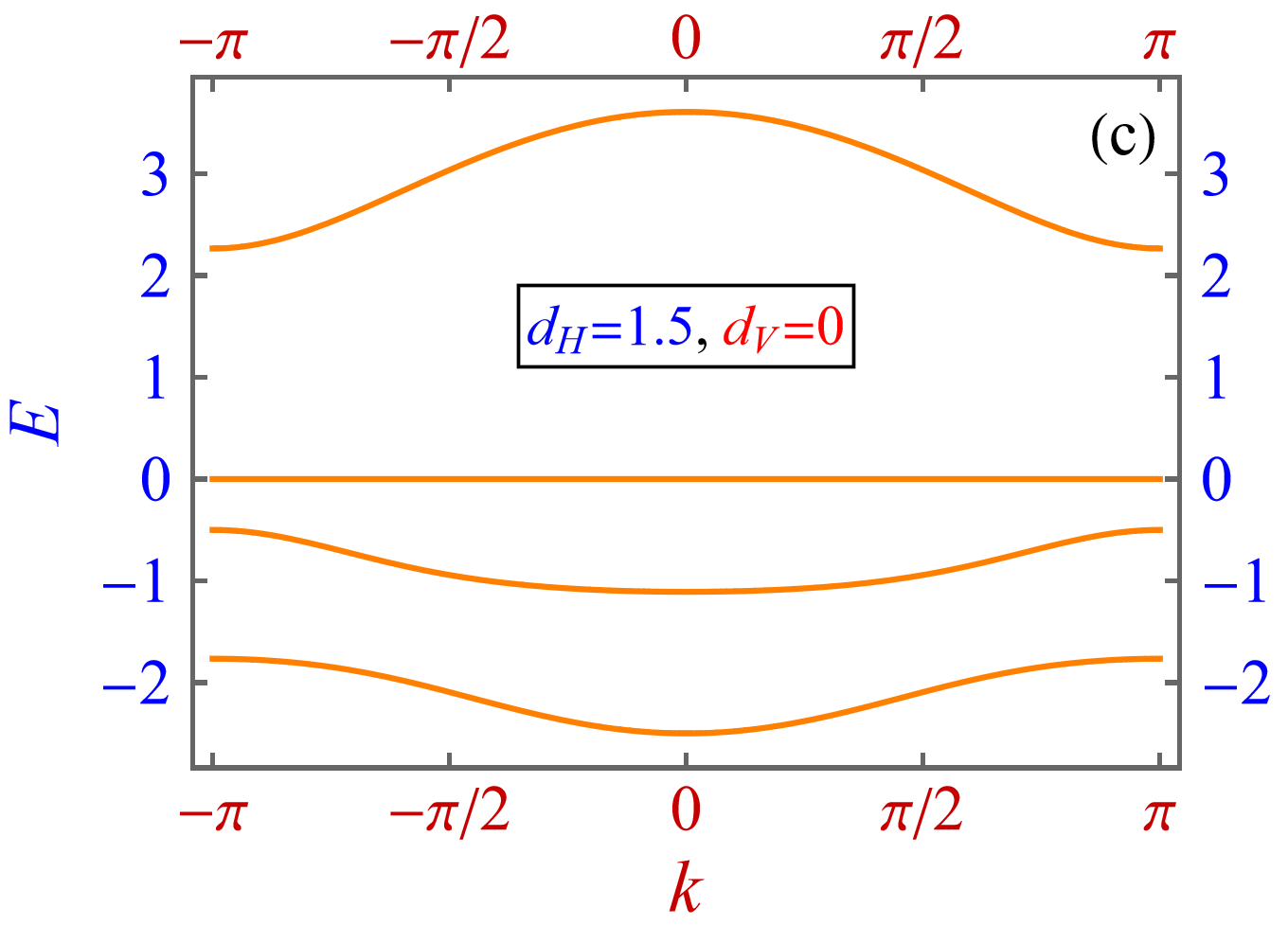}
\includegraphics[clip, width=0.49\columnwidth]{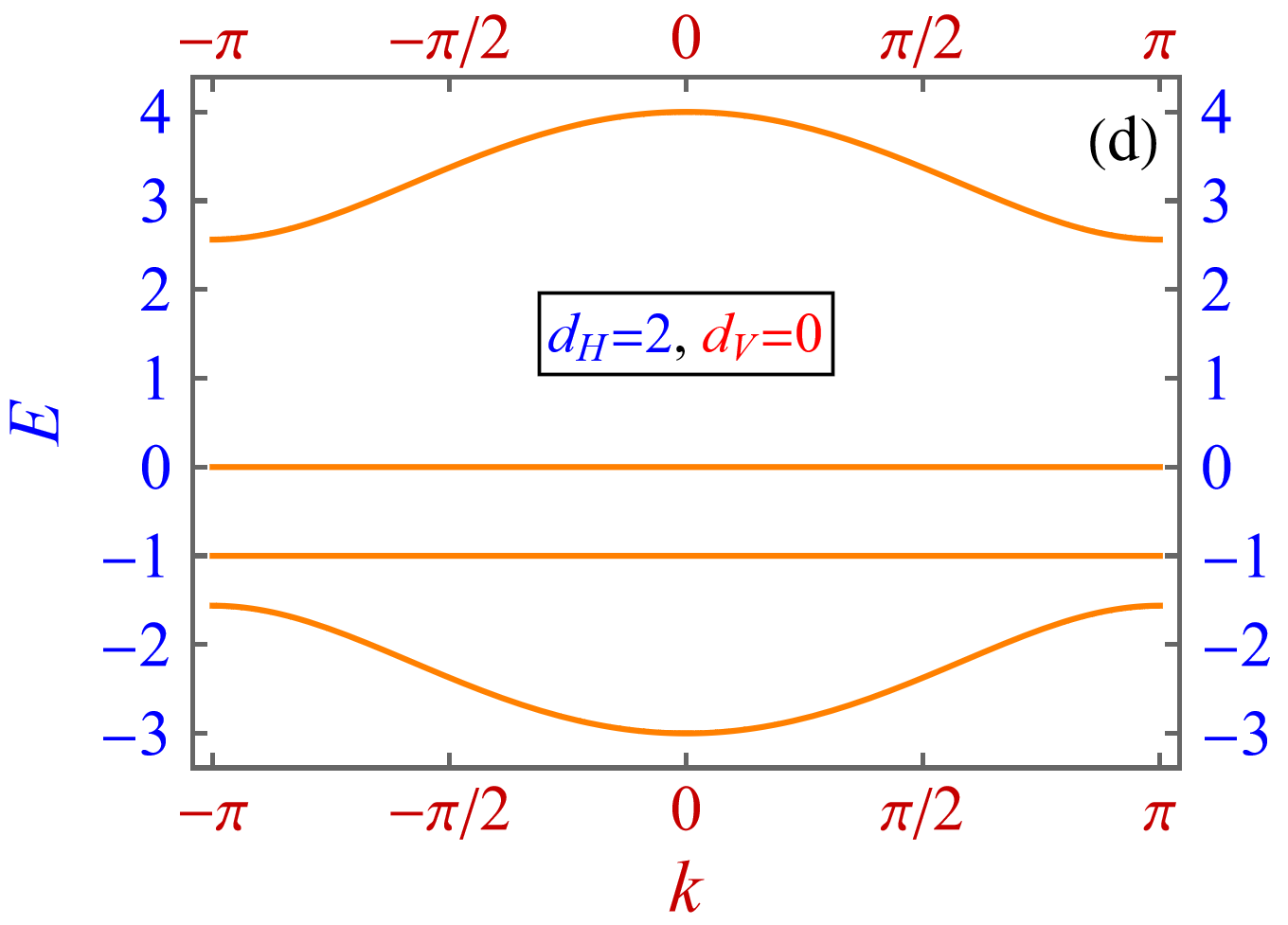}
\caption{Electronic band structure of the decorated diamond lattice model. 
(a) A gapless FB appears at $E=0$ for $d_H=0$, (b) one gapless and one gapped 
FB appear at $E=0$ and $E=-2$, respectively, for $d_H=1$, (c) the FB at $E=0$ is 
gapped out for $d_H=1.5$, (d) two gapped FBs emerge at $E=0$ and $E=-1$, 
respectively, for $d_H=2$. We have set $d_V=0$ and $\lambda=t=1$ in all the 
four cases.}
\label{fig:Band-structures}
\end{figure}
%
\subsection{Electronic band structure}
\label{subsec:band-structure}
The Hamiltonian of a quantum system contains nearly all the crucial information about the 
system, making it an extremely useful tool for analyzing the electronic properties of the system. 
In order to extract the information about the electronic band structure of this lattice model, 
we diagonalize the Hamiltonian in Eq.~\eqref{eq:hamilton-bloch}. We keep the values of the 
hopping parameters $\lambda$ and $t$ unaltered, and tune the $d_H$ and $d_V$ couplings in the 
system to examine what happens to the band structure. First, we tune the value of $d_H$ 
(measured in units of $t$) by keeping $d_V=0$. In a real system, such an effect can be 
achieved either by distorting the lattice by means of applying precisely controlled pressure 
or using two different types of chemical species for the top-bottom and the horizontal atomic sites. 
It is to be noted that, to physically realize the above control mechanism 
may be a nontrivial task, but not completely impossible. For this particular lattice model, one 
can apply the mechanical strain in a controlled fashion in between the two horizontal diagonal atomic 
sites to tune the horizontal diagonal hopping $d_H$; however, this will change the other hopping 
parameters like $t$ and $\lambda$. So, to keep $t$ and $\lambda$ unaltered, one has to tune the chemical 
potentials of the corresponding atomic sites by fine-tuning the gate-voltage~\cite{Gate-voltage-prl-2010}.
In a related context, it is worth to mention that, generation of pressure-induced flat bands in 
different lattice structure has been proposed in recent times~\cite{pressure-induced-FB-1,
pressure-induced-FB-2,pressure-induced-FB-3}.

For our lattice model, it is found that the number of FBs and whether they are gapped or gapless with 
other dispersive bands can be suitably controlled by tuning the value of $d_H$. The results are shown 
in Fig.~\ref{fig:Band-structures}. Note that, the energy ($E$) is measured in units of $t$. 
It is worth mentioning that, this model always provides a very robust FB at the energy $E=0$ 
irrespective of the value of $d_H$. However, by tuning $d_H$, one can generate additional FB 
in the energy spectrum, as well as it can be gapped out from the other dispersive bands in the 
energy spectrum. The opening of the band gap in the energy spectrum is also useful to analyze 
the topological properties of this lattice model, as discussed in Sec.~\ref{sec:topo-properties}. 

We note that, in this section, we have discussed the effect of tuning $d_H$ by keeping $d_V=0$. 
This makes sense, because as we increase the value of $d_H$, the coupling between the horizontal 
diagonal sites increases, which effectively means that the coupling between the vertical diagonal 
sites will vanish. However, it is also interesting to look into the opposite scenario, where we 
make the value of $d_V$ nonzero by keeping $d_H=0$. This case is discussed in the 
Appendix~\ref{sec:appendix-1}. 
\subsection{Compact localized states}
\label{subsec:CLS}
%
\begin{figure}[ht]
\centering
\includegraphics[clip, width=0.195\columnwidth]{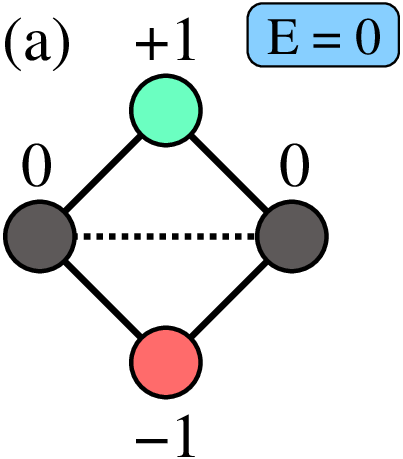}
\includegraphics[clip, width=0.38\columnwidth]{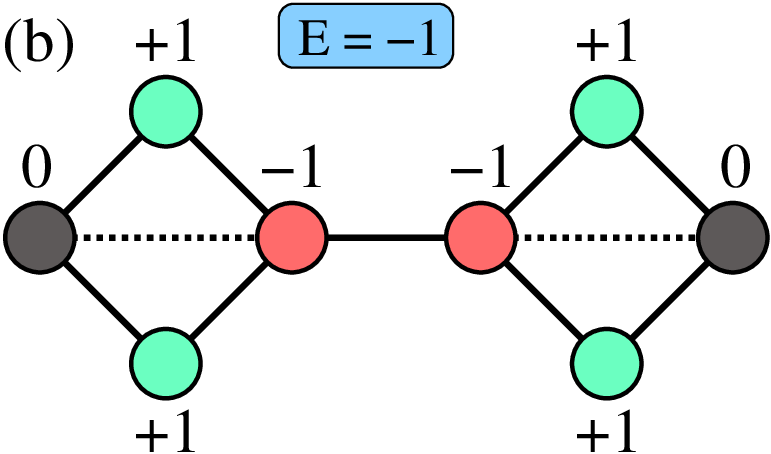}
\includegraphics[clip, width=0.38\columnwidth]{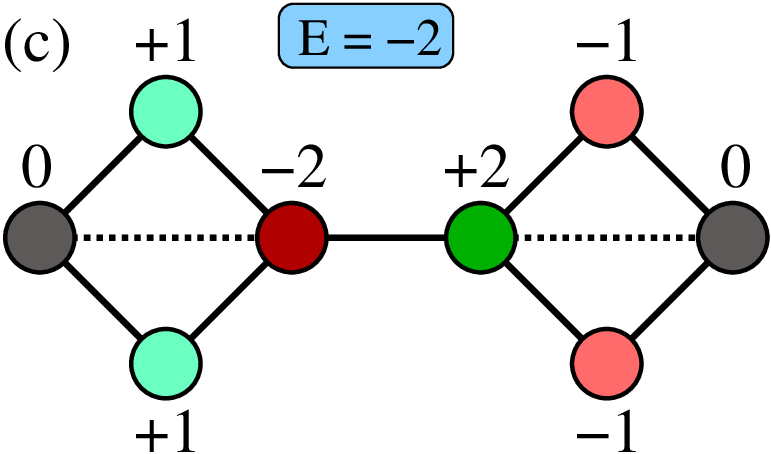}
\caption{The real-space wave function amplitude distribution corresponding 
to the flat bands at the energies 
(a) $E=0$ (for all possible values of $d_H$), 
(b) $E=-1$ (for $d_H=2$), and 
(c) $E=-2$ (for $d_H=1$). Zero 
wave function amplitudes are denoted by grey-shaded circles, while the 
nonzero wave function amplitudes are marked by green (for positive amplitudes) 
and red (for negative amplitudes) circles, respectively. 
We have taken $d_V=0$ and $\lambda=t=1$ in all the three cases.}
\label{fig:CLS}
\end{figure}
%
The appearance of the FBs in the energy spectrum elaborated in subsection~\ref{subsec:band-structure} 
is a momentum ($k$)-space description. In the present subsection, we will discuss how one can 
classify such FB states using a real space description. It is in general true that, corresponding 
to the FB energies in the electronic band structure of a lattice model, it is always possible to 
construct a special kind of eigenstate in the real space in which we can have nonzero wave function 
amplitude distribution only over a few unit cells of the lattice, and they are connected with each 
other with strictly zero wave function amplitudes. Such real-space configurations of the FB states 
are known as the compact localized states. The construction of the CLS can be imagined to act 
as a \emph{quantum prison}, in which the electrons are completely localized at the FB energies even 
in the absence of any disorder in the system. 

For the decorated diamond lattice model, the construction of the CLS corresponding to the various 
FB energies is exhibited in Fig.~\ref{fig:CLS}. We observe that for the FB at $E=0$, the CLS spans 
over only one unit cell, whereas for the FBs at $E=-1$ and $E=-2$, the CLS are spread over two unit 
cells. Such spread of the CLS is often connected to a boost of the quantum metric, which is the 
real part of the quantum geometric tensor, and may be useful in the study of unconventional FB 
superconductivity~\cite{Quantum-metric-prb-2022}. The highly localized character of the eigenstates 
corresponding to the CLS will be more clear to us from the analysis of the density of states of the 
system, which is discussed in the following subsection. 
Before we end this subsection, it is worthwhile to highlight an interesting 
remark: in some cases, CLS may not be the complete basis functions of the FB; one may construct 
certain noncontractible loop states which must come from another dispersive band. Hence, one 
of the dispersive bands touches the flat band at exactly 
one point~\cite{Balents-prb-2008,Bernevig-prl-2020}.
\subsection{The density of states}
\label{subsec:DOS}
%
\begin{figure}[ht]
\centering
\includegraphics[clip, width=0.49\columnwidth]{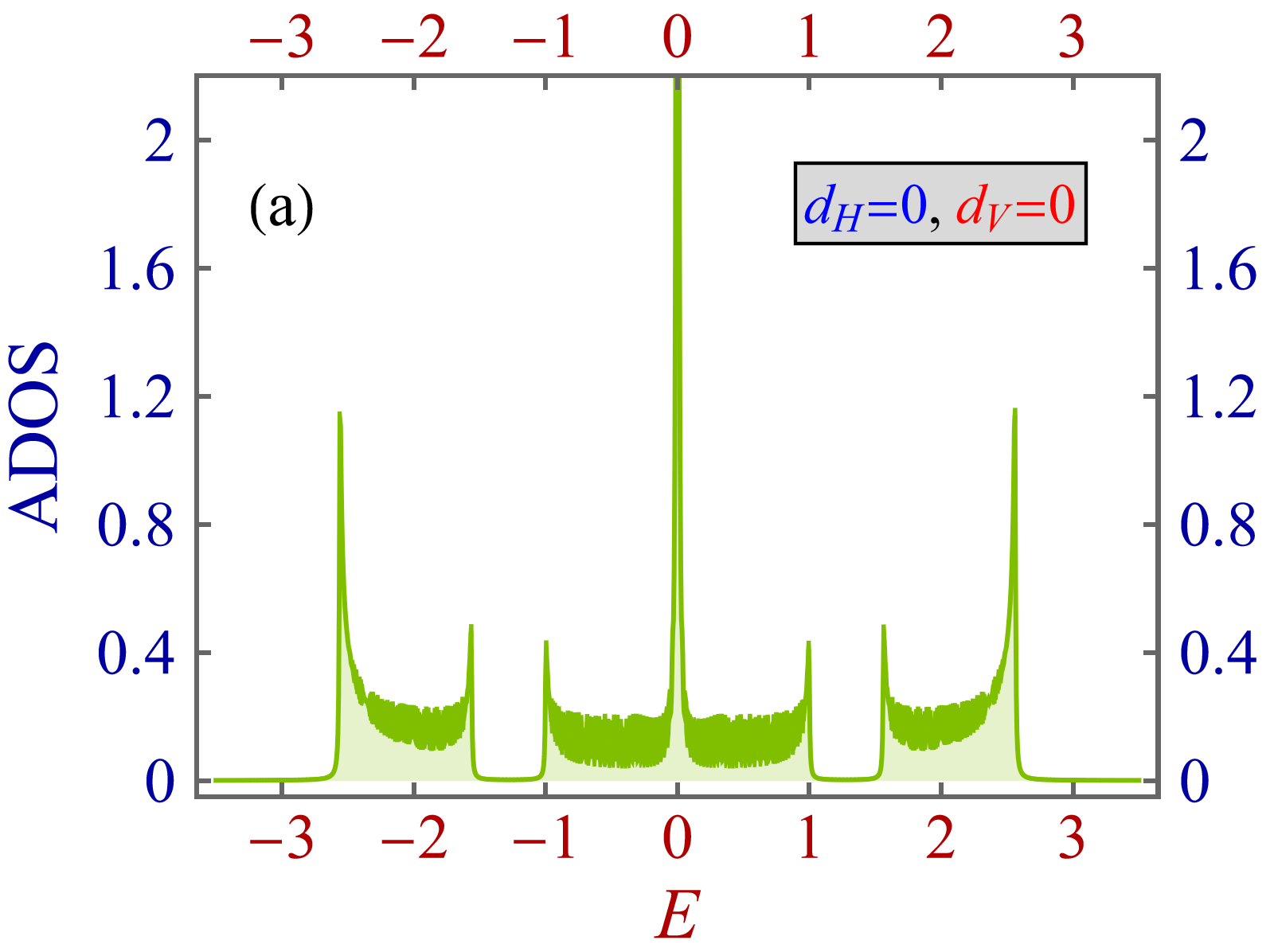}
\includegraphics[clip, width=0.49\columnwidth]{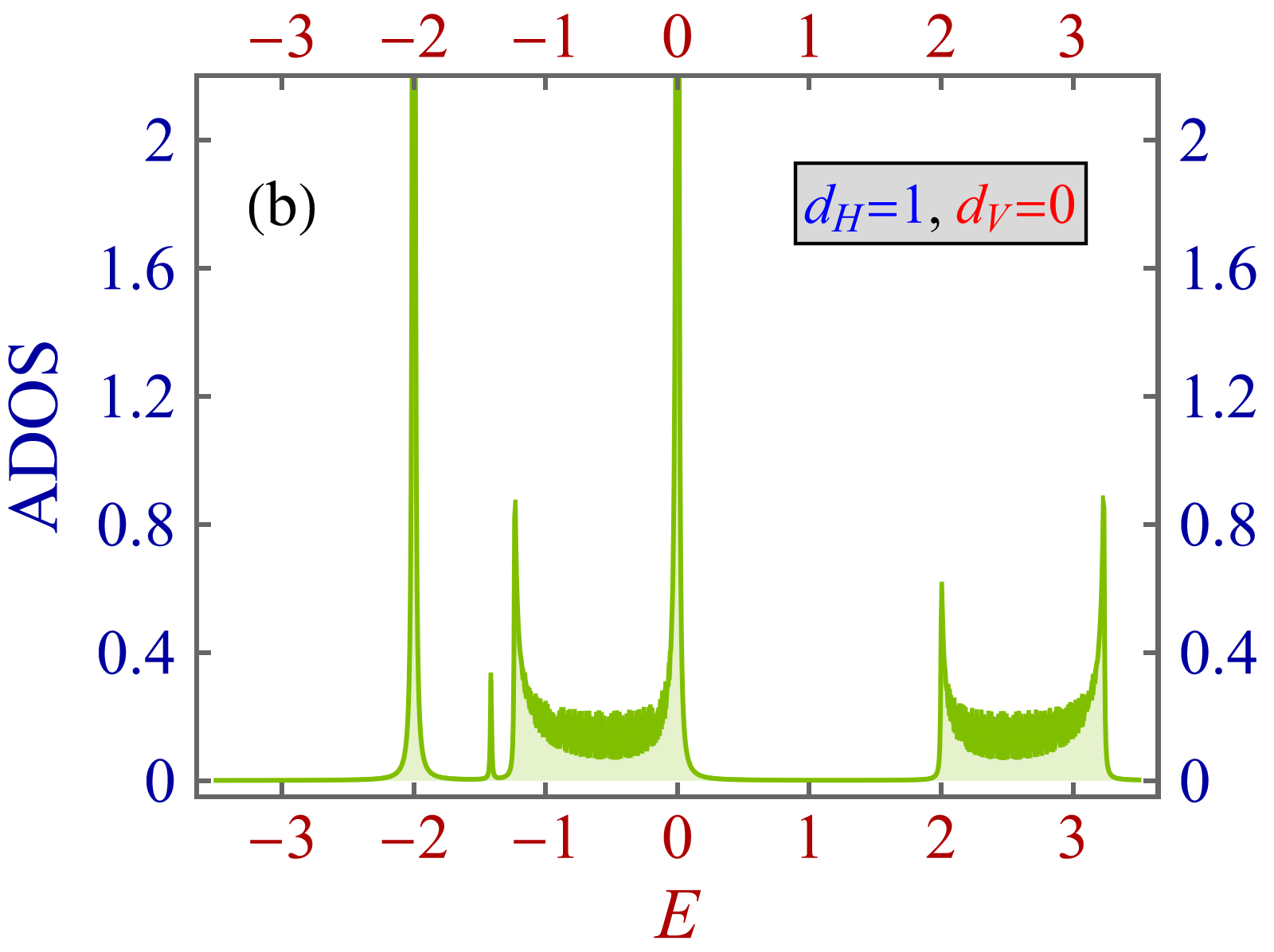}
\vskip 0.4cm
\includegraphics[clip, width=0.49\columnwidth]{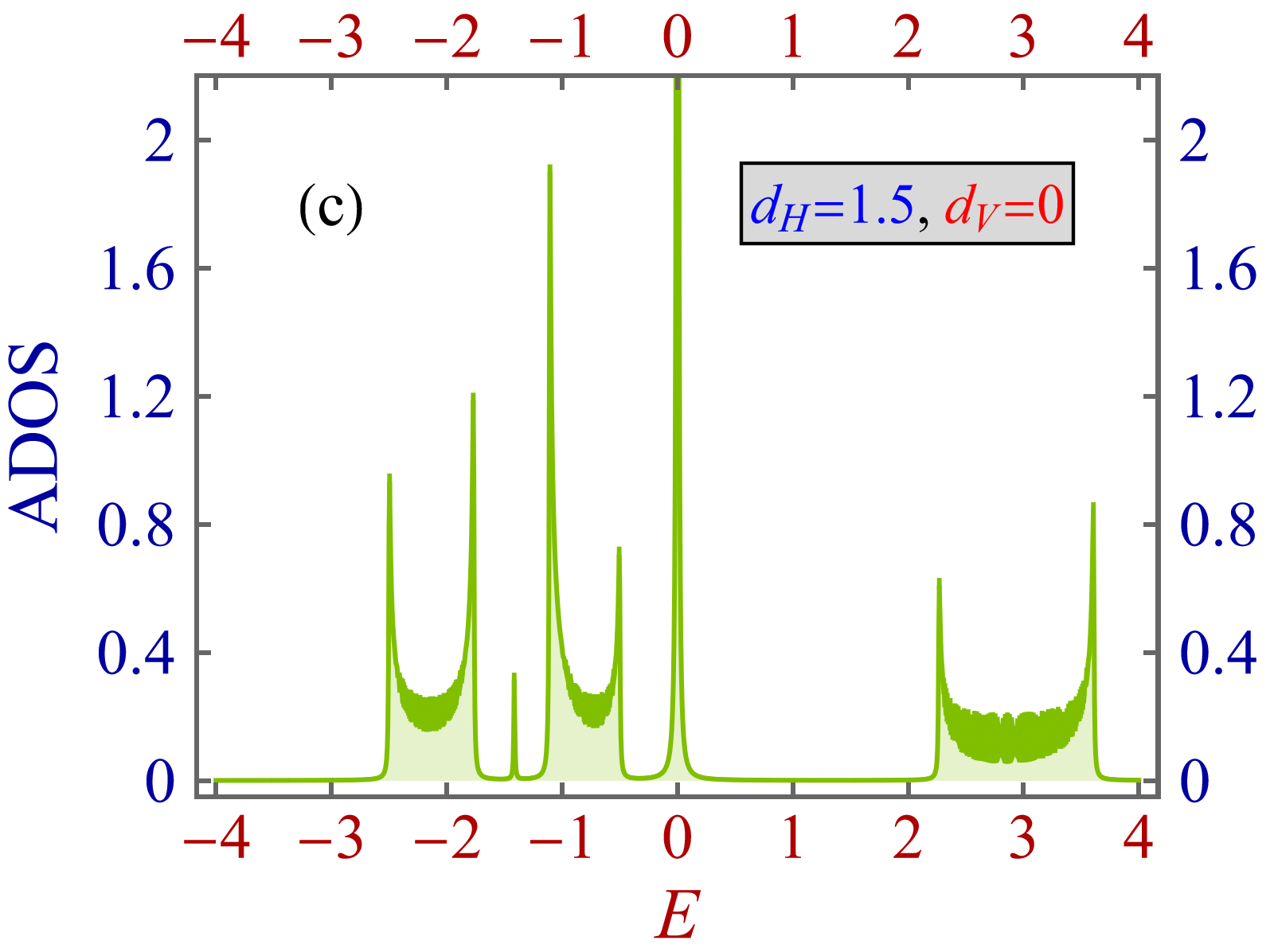}
\includegraphics[clip, width=0.49\columnwidth]{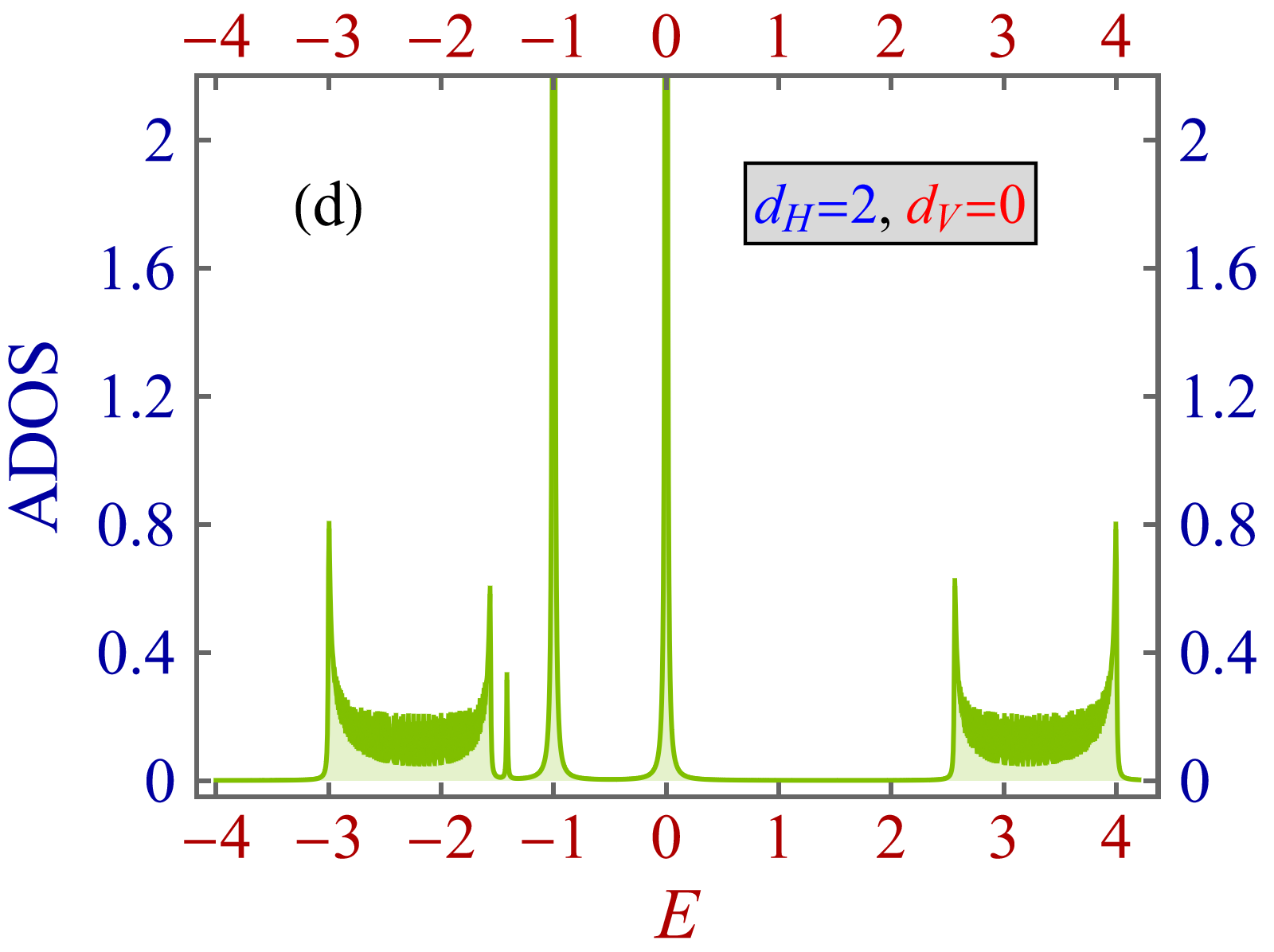}
\caption{Variation of the average density of states (ADOS) as a function of the 
energy ($E$) for the decorated diamond chain. The total number of unit cells 
considered here is $100$ (\emph{i.e.}, $\mathcal{N} =400$ sites). The plots are done for different 
values of the horizontal diagonal coupling $d_H$ (measured in units of $t$), viz., 
(a) $d_H=0$, (b) $d_H=1$, (c) $d_H=1.5$, and (d) $d_H=2$. The other parameters are chosen 
as $d_V=0$ and $\lambda=t=1$ in all four cases. The value of the small imaginary part 
$\eta$ is taken to be equal to 0.005.}
\label{fig:DOS}
\end{figure}
%
The highly localized character of the FB states in a lattice model can be distinctly identified by 
looking at its density of states spectrum. One can easily compute the average density of states (ADOS) 
of a finite-size system by using the Green's function formalism given by,   
\begin{equation}
\rho(E) = -\dfrac{1}{\mathcal{N} \pi}\textrm{Im} 
\Big[ \textrm{Tr}\left[\bm{\mathcal{G}}(E)\right]\Big],
\label{eq:ADOS}
\end{equation} 
where $\bm{\mathcal{G}}(E) = \big[(E + i\eta){\bm I} - 
\bm{H} \big]^{-1}$ (with $\eta \rightarrow 0^{+}$), ${\bm I}$ is an identity matrix of $\mathcal{N} \times \mathcal{N} $ 
dimension, $\mathcal{N}$ being the total number of sites in the lattice, and `$\textrm{Tr}$' denotes the trace 
of the Green's function matrix $\bm{\mathcal{G}}$. By using the above formula, we have computed the 
ADOS for a finite-size decorated diamond chain with $100$ unit cells (\emph{i.e.}, $\mathcal{N}=400$ sites). 
The results are displayed in Fig.~\ref{fig:DOS}. From Fig.~\ref{fig:DOS}, we can clearly identify 
the appearance of the sharp peaks in the ADOS spectrum corresponding to the highly localized states 
exactly at the FB energies. The other relatively smaller picks indicate the band edges. From the 
ADOS spectrum, one can clearly distinguish whether the FBs are gapped or gapless from the other 
dispersive continuum bands. We note that the ADOS spectrum exactly corroborates the information 
we have obtained about the FBs in Fig.~\ref{fig:Band-structures}. 
\subsection{The robustness of the FBs against disorder}
\label{subsec:disorde-effect}
%
\begin{figure}[ht]
\centering
\includegraphics[clip, width=0.49\columnwidth]{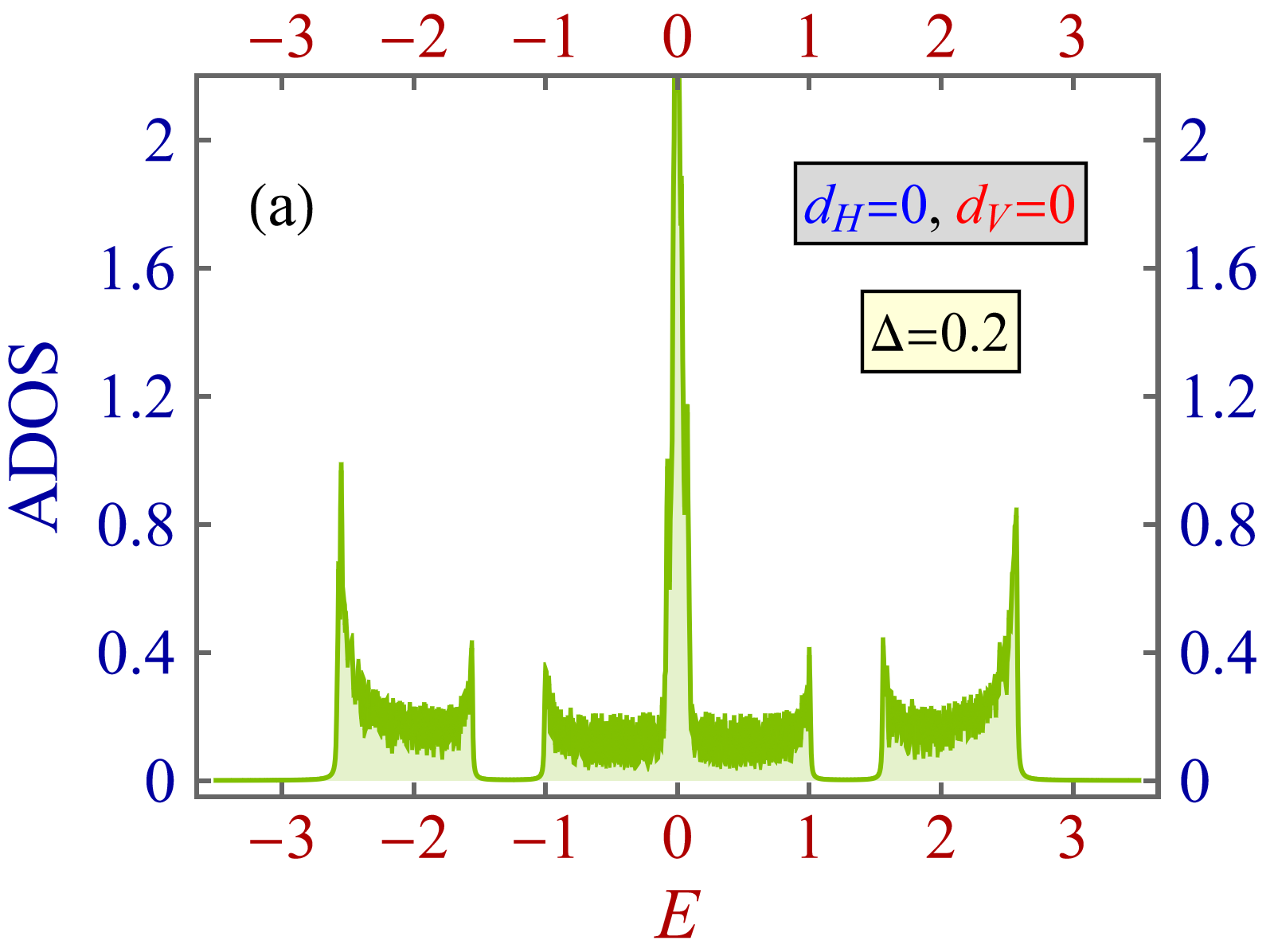}
\includegraphics[clip, width=0.49\columnwidth]{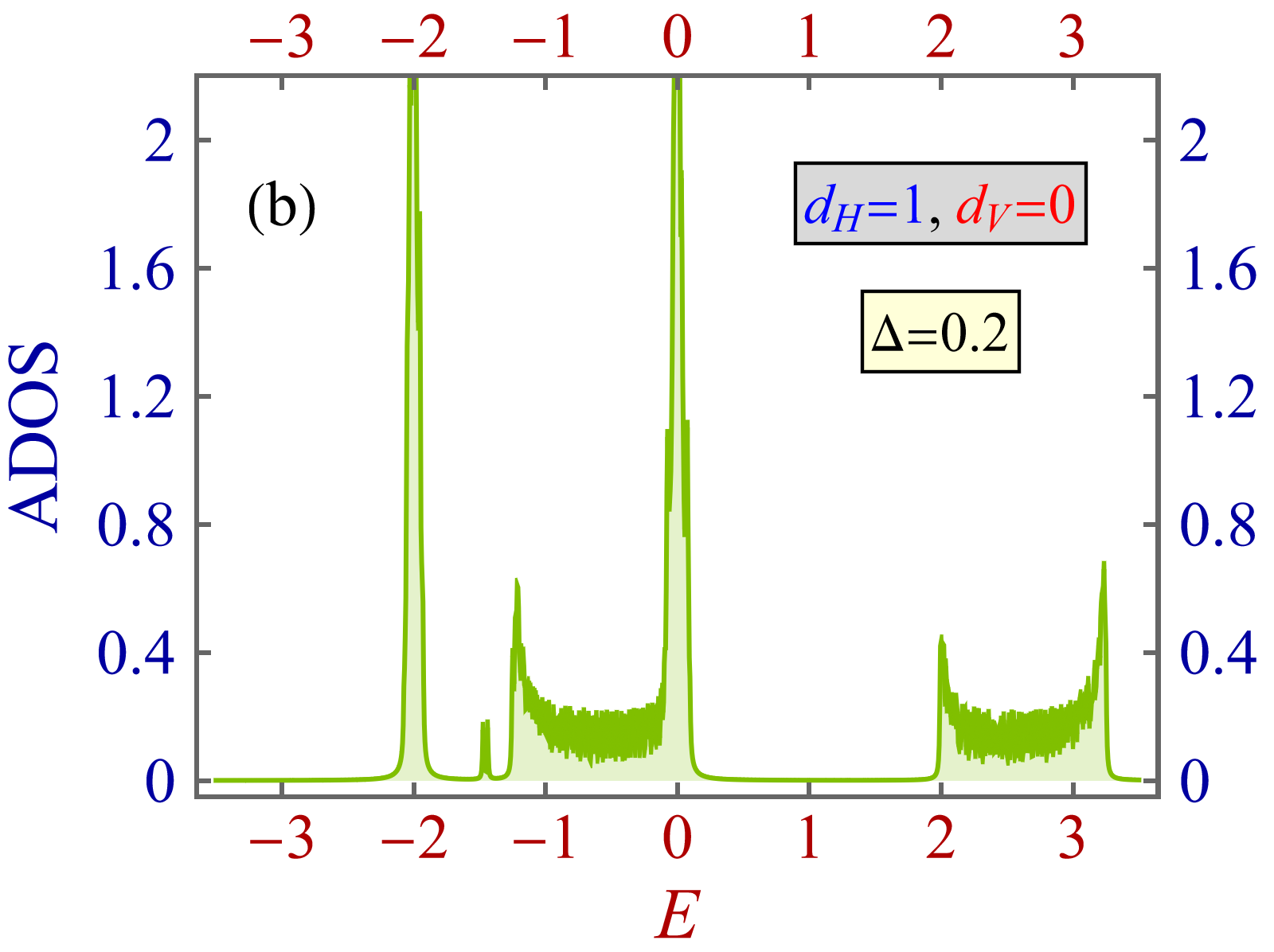}
\vskip 0.4cm
\includegraphics[clip, width=0.49\columnwidth]{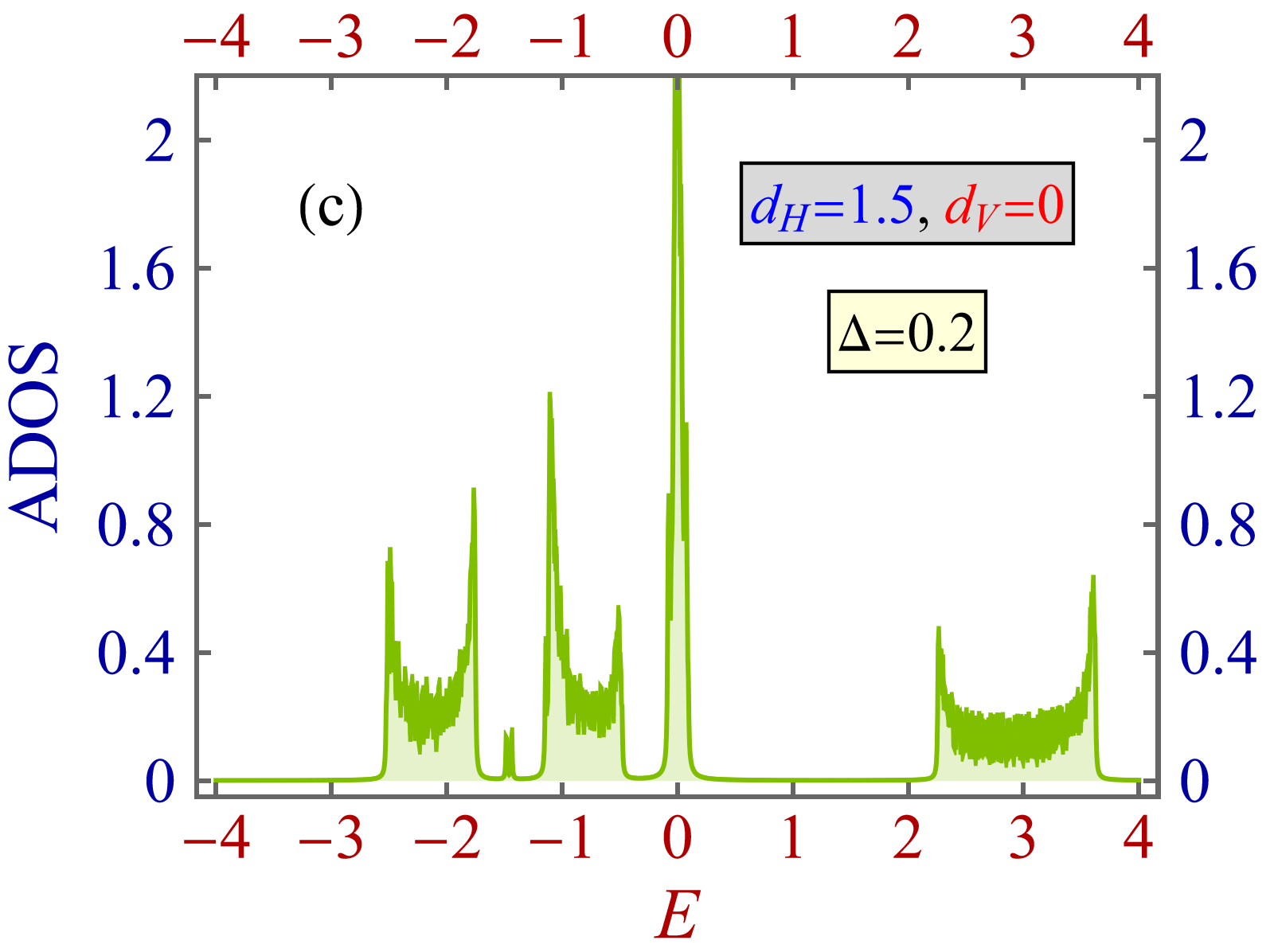}
\includegraphics[clip, width=0.49\columnwidth]{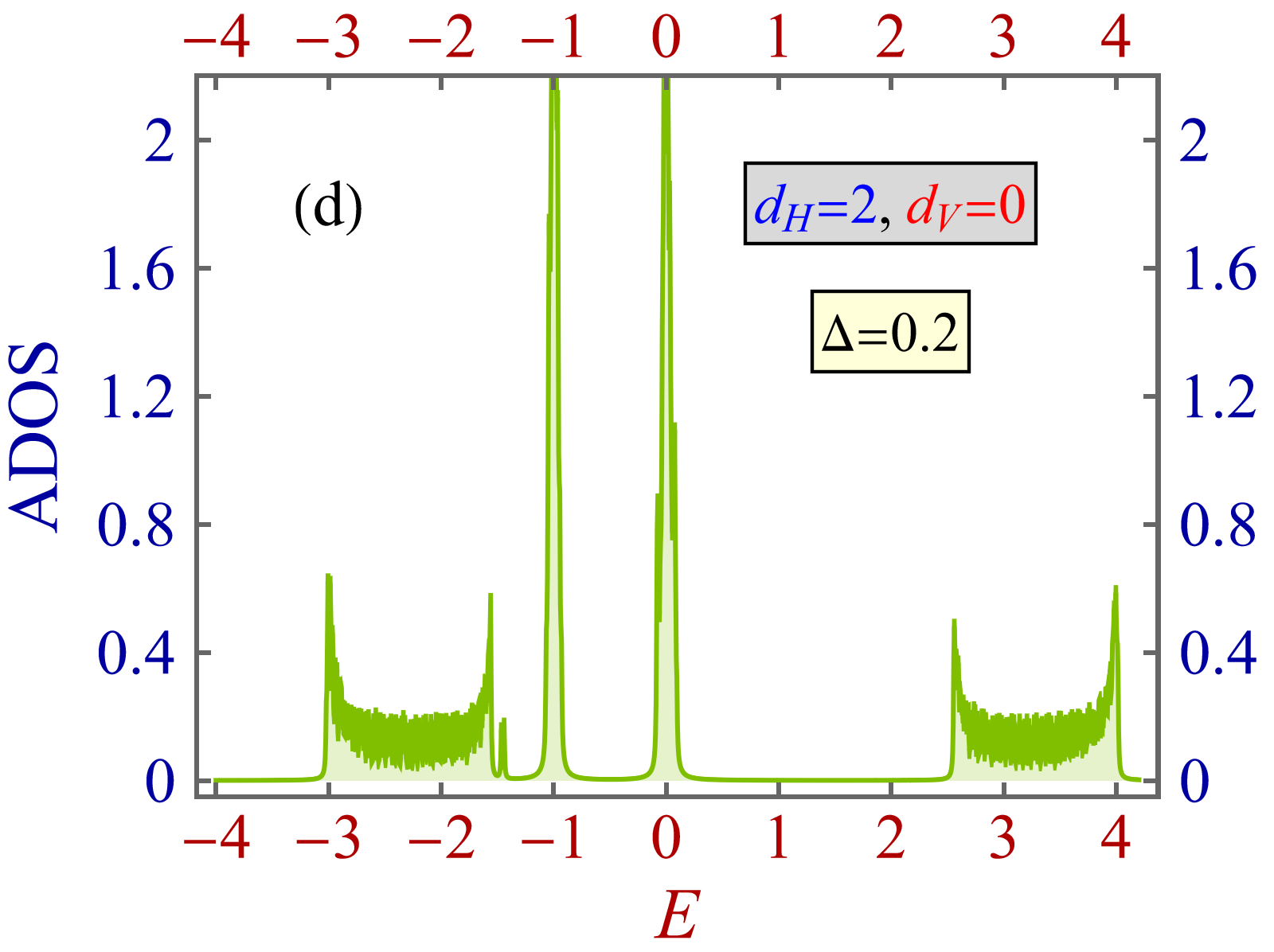}
\caption{Variation of the average density of states (ADOS) as a function of the 
energy ($E$) for the decorated diamond chain by incorporating a random onsite 
(diagonal) disorder of strength $\Delta=0.2$ (measured in units of $t$) in the system. 
All other parameters are kept unaltered as in Fig.~\ref{fig:DOS}.}
\label{fig:DOS-with-small-disorder}
\end{figure}
%
It is very important to examine the robustness of these FB states against the small amounts of 
disorder in the system. In this subsection, we discuss this aspect. In general, when a ﬂat band 
is separated from the other dispersive bands in the spectrum by a gap, the compact localized states 
are robust against the disorder, retaining their strong localization character under the ﬁnite 
disorder strength.  However, when the band gap vanishes, the eﬀective disorder potential is enhanced. 
In that scenario, the compact localized states are destroyed, and they acquire large localization 
lengths with different power law scalings as compared to the Anderson localization length. 

For our lattice model, we introduce a random diagonal disorder in the onsite potential distribution 
of the system $\varepsilon_{i}\sim[-\frac{\Delta}{2},\frac{\Delta}{2}]$, where $\Delta$ denotes the 
disorder strength. We find that, for a relatively small disorder strength $\Delta=0.2$ (measured in 
units of $t$), the highly localized FB states in the ADOS spectrum remain intact as depicted in 
Fig.~\ref{fig:DOS-with-small-disorder}. However, as we increase the strength of the onsite disorder $\Delta$ 
to higher values, the highly localized FB states collapse, and we can have a localization-to-delocalization 
transition in the system known as the inverse Anderson 
transition~\cite{Inverse-Anderson-transition-1,Inverse-Anderson-transition-2,Inverse-Anderson-transition-3}. 
In order to substantiate this claim, we have computed the ADOS for a higher disorder strength 
$\Delta=0.8$ (measured in units of $t$), where the FB peaks in the spectrum begin to collapse. We have also 
calculated the inverse participation ratio (IPR) to quantitatively understand the character of these localized 
states. We append these results in the Appendix~\ref{sec:appendix-2}. 
\section{Topological properties of the model}
\label{sec:topo-properties} 
In the present section, we present the results related to the topological properties offered by the 
decorated diamond lattice model. Comprehending the topological characteristics of a lattice model 
often has important ramifications for emerging quantum technologies. For our lattice model, the 
topological nature of the bands is characterized by the Zak phase or, equivalently, by the winding 
number. This topological invariant corresponds to the Berry phase~\cite{berry-1984}, which can be 
expressed in terms of the $n$-th Bloch eigenstates as 
\begin{equation}
\nu = -\dfrac{i}{\pi} \oint \langle u_{k,n}| \dfrac{du_{k,n}}{dk} \rangle dk.
\label{eq:winding}
\end{equation}
The Zak phase is related to the winding number by $\mathcal{Z} = \pi \nu$. Experimentally, the 
topological signature of a system has been demonstrated through the measurement of the winding number 
in analogous photonic networks~\cite{exp-winding-2018,exp-winding-2020}. In our analysis, we compute 
the winding number using the gauge-invariant Wilson loop method~\cite{zak-fukui-2005}. According to 
Eq.~\eqref{eq:winding}, the integration is converted into the summation over the entire Brillouin Zone 
by dividing it into $N$ number of similar discrete sections, which ensures that each interval in the 
wave vector will be $\delta k = 2\pi/Na$ with lattice constant $a$. For our computation, we consider 
$N=401$, which provides a sufficiently accurate approximation of the integration. Within the Wilson loop 
framework, for a non-degenerate $n$-th band, the winding number is given 
by~\cite{zak-fukui-2005, zak-marzari-1997}, 
\begin{equation}
\mathcal{\nu} = -\dfrac{1}{\pi} \textrm{Im} \bigg[ \log \Big(\prod_{k_{j}} \langle u_{k_{j},n}| u_{k_{j+1},n} \rangle \Big)\bigg].
\label{eq:winding-Wilson-loop}
\end{equation}
It is worthwhile mentioning that, here we are calculating the single-band winding number. 
If two (or more) bands touch each other, then the single-band winding number becomes 
ill-defined~\cite{winding-close,sheet} at those band touching points. Even in those cases also, one can 
still evaluate the multi-band topological invariants by grouping the bands, or may compute the winding 
around the band touching point. However, for our analysis, we are only focusing on the single-band winding 
number over the first Brillouin zone when all the bands are separated by band gaps.
%
\begin{figure}[ht]
\centering
\includegraphics[clip, width=0.49\columnwidth]{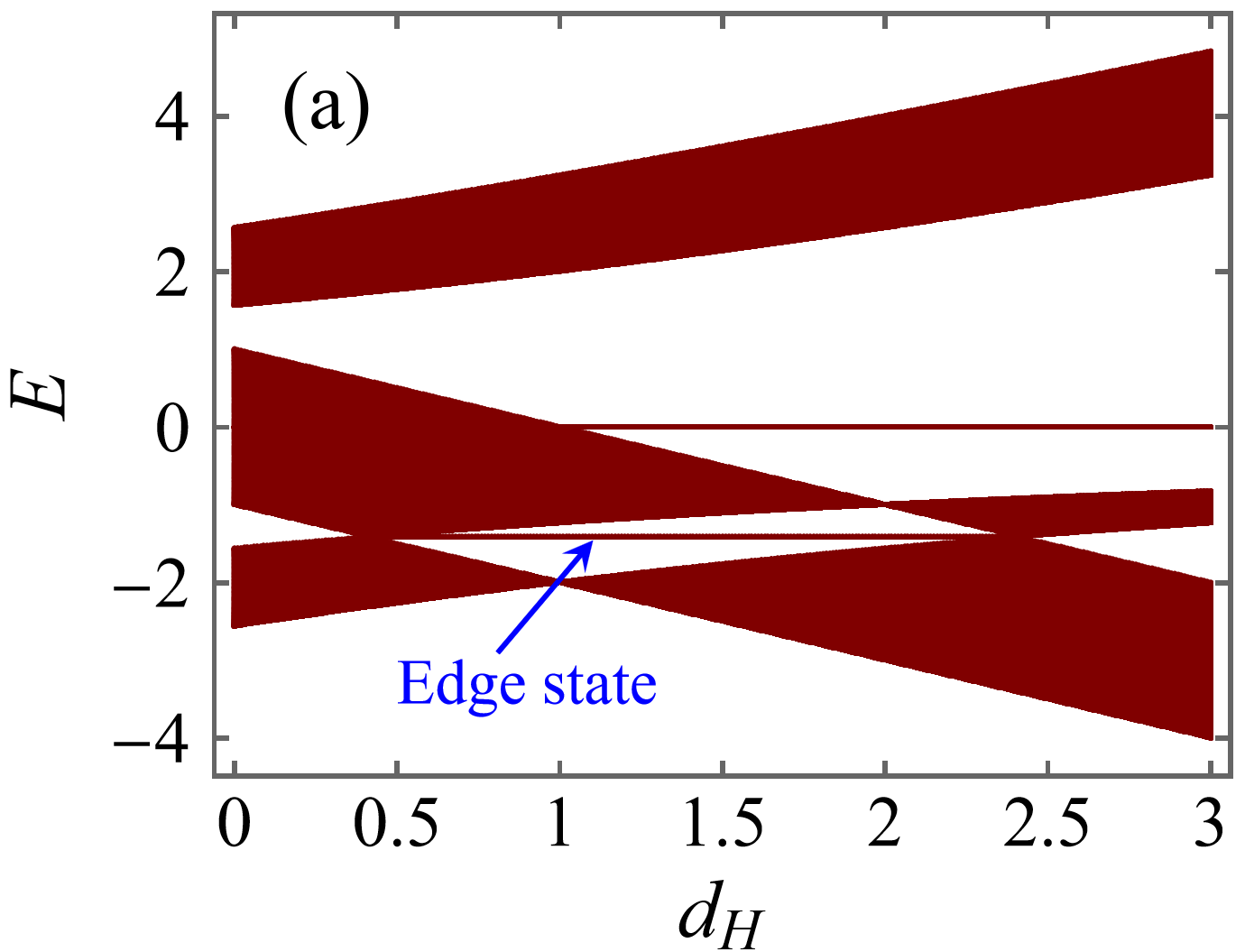}
\includegraphics[clip, width=0.49\columnwidth]{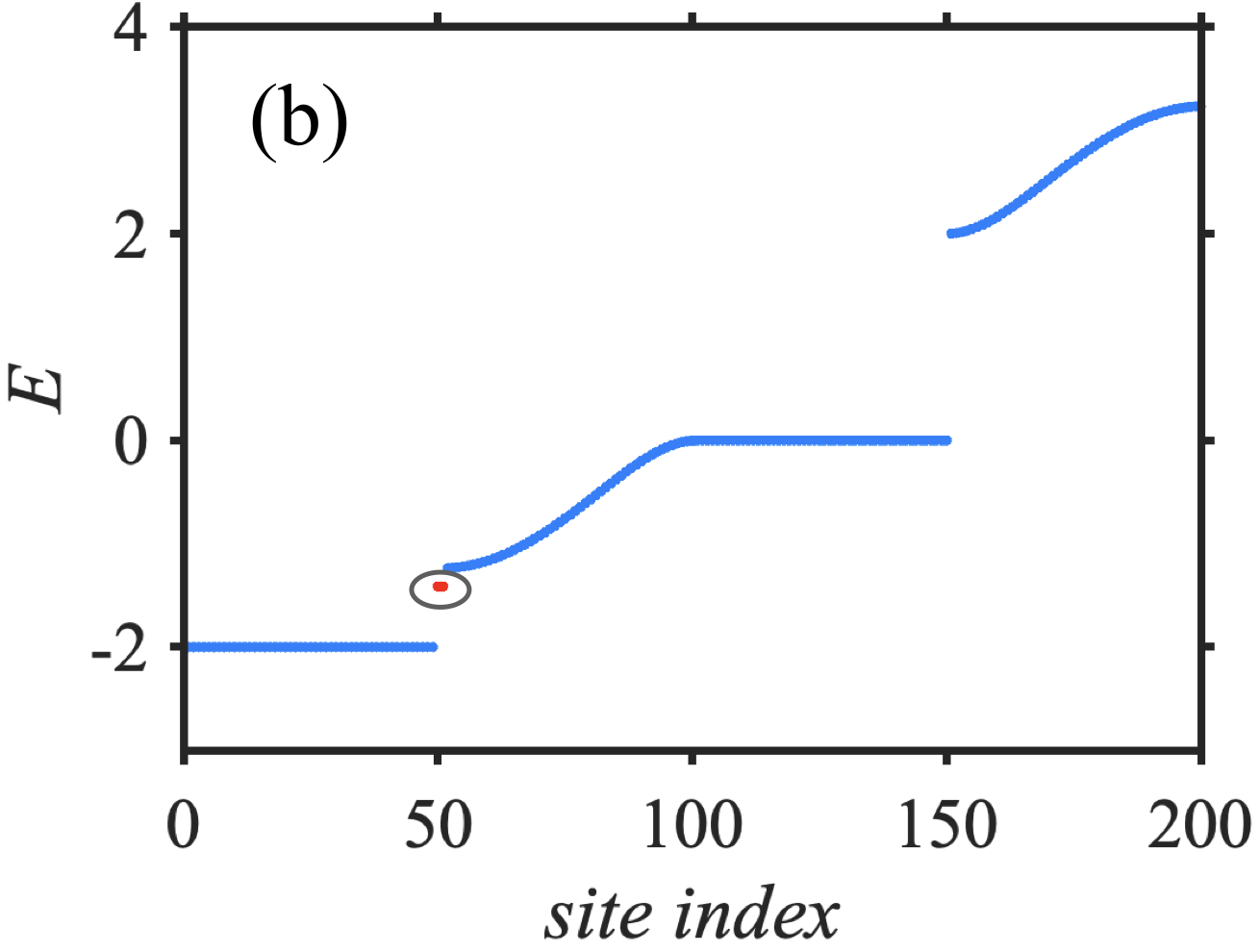}\\
\vskip 0.4cm
\includegraphics[clip, width=0.49\columnwidth]{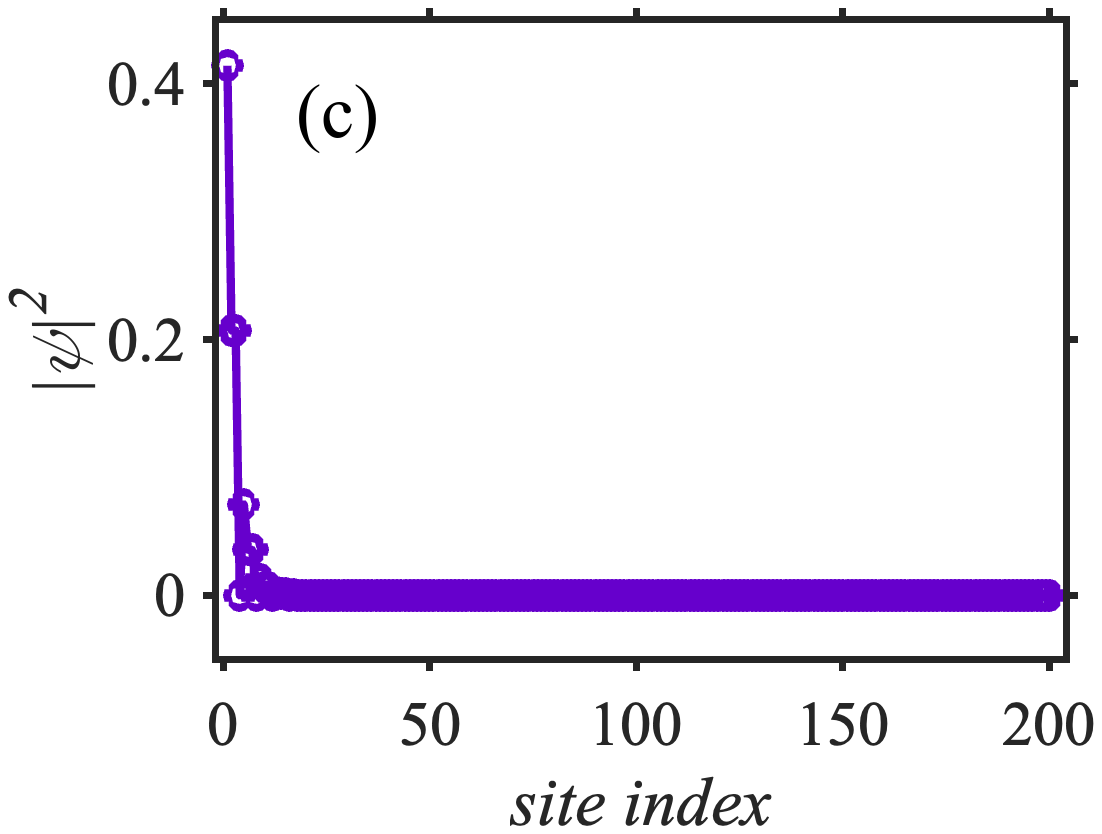}
\includegraphics[clip, width=0.49\columnwidth]{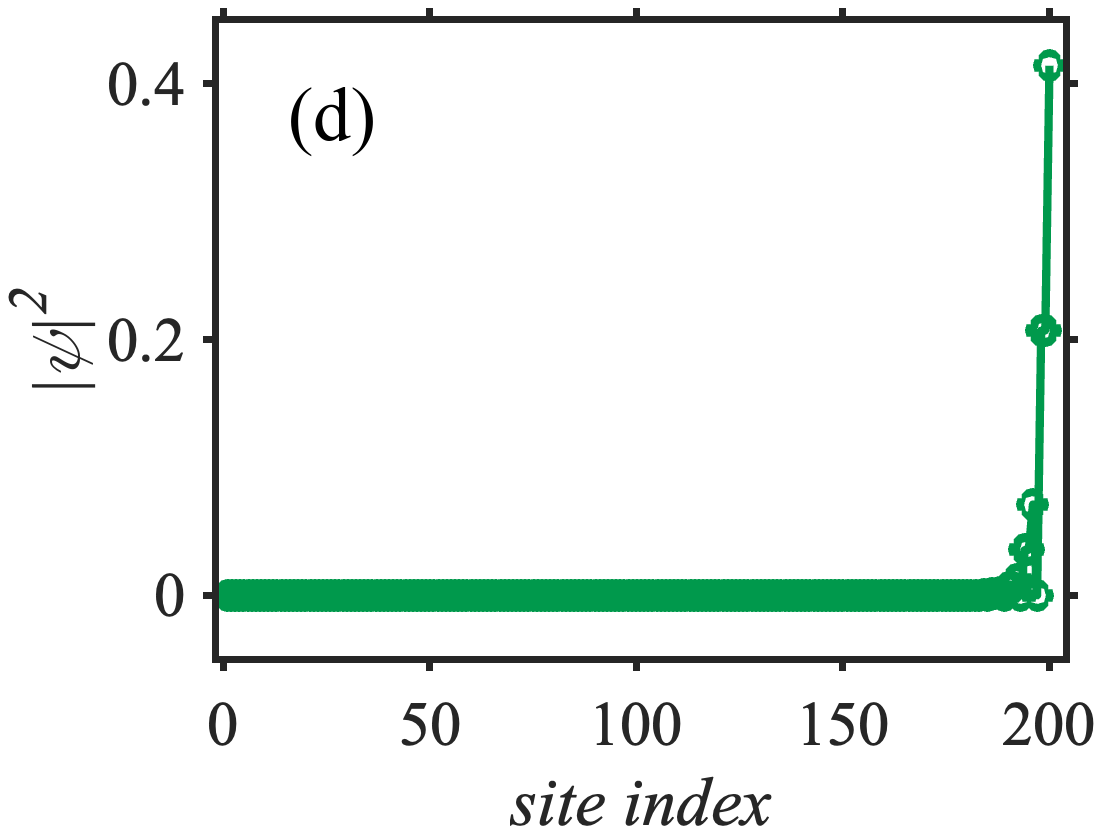}
\caption{(a) Energy eigenspectra of the lattice model as a function of the 
\emph{horizontal} diagonal coupling $d_{H}$ to mark the emergence of the 
edge states by tuning $d_{H}$. The edge states are two-fold degenerate 
and appear after a critical value of $d_H$ at the energy $E=-\sqrt{2}$. 
(b) Distribution of the energy eigenspectra for $d_{H}=t$ with an open 
boundary condition for $50$ unit cells. The in-gap degenerate edge states 
are marked with an encircled red dot at the energy eigenvalue $E=-\sqrt{2}$. 
(c, d) Amplitude distributions of one pair of in-gap edge states with energy 
$E=-\sqrt{2}$ for $50$ unit cells with an open boundary condition. 
The distribution for one state is confined to the left edge 
(shown in violet color), while the other one is strictly localized at the 
right edge (shown in green color). The other parameters are fixed at 
$\lambda=t=1$ and $d_{V}=0$.}
\label{fig:topology}
\end{figure}
%

Till now, we have discussed the quasi-one-dimensional decorated diamond lattice with periodic boundary 
conditions. Now, we focus on the scenario with open boundary condition to explain the emergence of 
topological nontrivial features for this lattice model as exhibited in Fig.~\ref{fig:topology}. 
Fig.~\ref{fig:topology}(a) explains the energy spectrum of the system as a function of the 
horizontal diagonal coupling parameter $d_H$. Beyond a critical value of $d_H$, 
we observe the emergence of in-gap edge states, which persist over a finite range of $d_H$. 
Within this regime, the bulk bands become gapped after a band-closing transition, indicating a 
topological phase transition into a nontrivial phase. These edge states disappear upon imposing 
periodic boundary conditions, confirming their boundary-localized nature. In particular, 
Fig.~\ref{fig:topology}(a) reveals that the edge states are pinned at energy $E = -\sqrt{2}$ for a 
specific range of $d_H$. Throughout these calculations, we fix other parameters $\lambda=t=1$ and $d_{V}=0$. 
It is to be noted that, the flat line appearing at the energy $E=0$ in 
Fig.~\ref{fig:topology}(a) is the FB state, not an edge state, as already shown earlier in 
Fig.~\ref{fig:Band-structures}.

\begin{figure}[ht]
\centering
\includegraphics[clip, width=0.5\columnwidth]{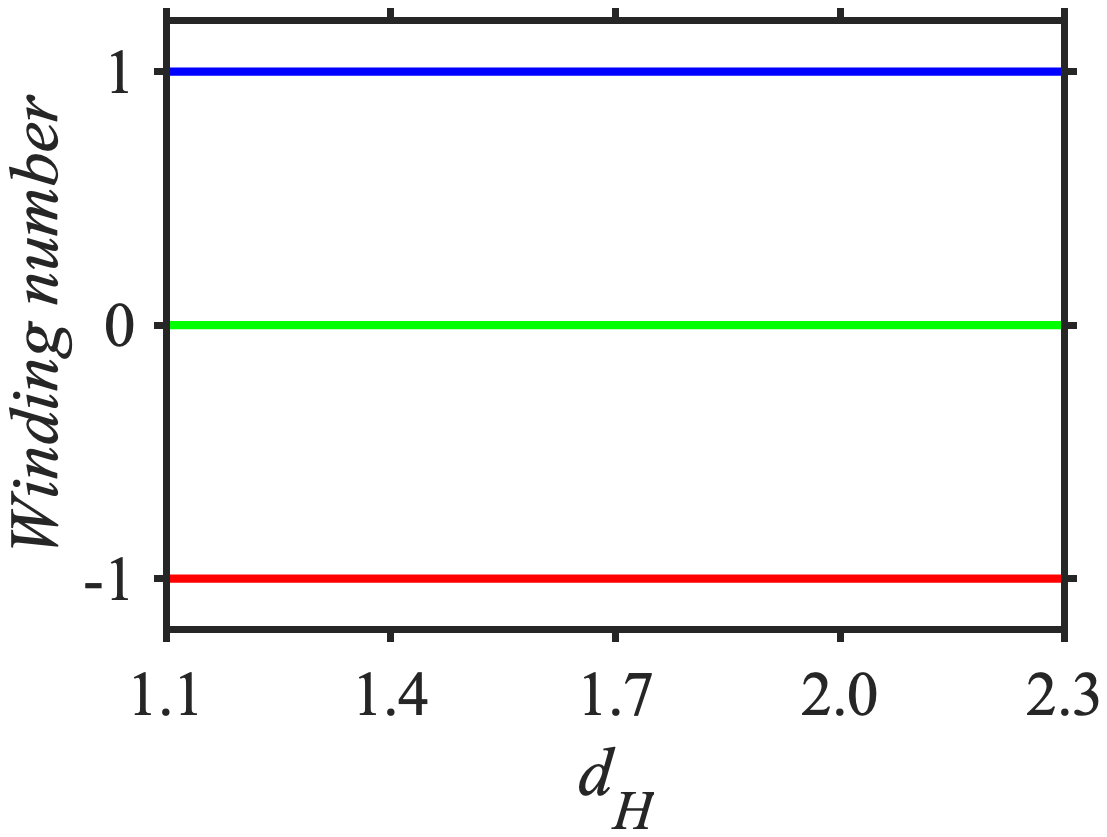}
\caption{The variation of the single-band winding number 
as a function of $d_{H}$ for all the four bands, which clearly highlights the 
topologically nontrivial regime. The lowest and the second lowest bands have 
the winding number $-1$ and $+1$, which are shown by the red and blue colors, 
respectively. The rest of the bands are topologically trivial with zero 
winding number, as shown by the green color.}
\label{fig:winding-number-distribution}
\end{figure}

To further explore this, we consider $d_H = t$ and plot the eigenvalues with respect to the site index for 
a system size with $50$ unit cells in Fig.~\ref{fig:topology}(b). The degenerate edge states are clearly 
identified and marked with an encircled red dot, showing a two-fold degeneracy. As depicted in 
Fig.~\ref{fig:topology}(c) and (d), one of these degenerate edge states is localized at the left edge of 
the system, while its partner is confined to the right edge of the system. These states are exponentially 
localized near the edges and decay very rapidly into the bulk, which is a hallmark of the topologically 
protected modes. To characterize the topological nature of this phase, we compute the single-band 
winding number as a function of $d_H$, which yields integer values within the nontrivial topological zone. This 
quantized invariant confirms the presence of topologically protected edge states through the principle of 
bulk-boundary correspondence. The single-band winding number distribution, shown 
in Fig.~\ref{fig:winding-number-distribution}, clearly distinguishes the topologically trivial and nontrivial 
phases, with nonzero integer values of winding numbers indicating topological behavior. 
%
%

We have analyzed the variation of the single-band winding number for all gapped bands to 
trace the evolution of topological characteristics in the system. The winding number is well-defined only 
when the corresponding band remains isolated by an energy gap from its neighboring bands. Beyond a critical 
value of $d_{H}$, edge states emerge as a gap opens between the lowest and the second bands. In this 
nontrivial topological regime, the winding number of the lowest band takes the value $-1$, while that of 
the second band becomes $+1$, as illustrated in Fig.~\ref{fig:winding-number-distribution} by the red and 
blue lines, respectively. In contrast, the third and the highest bands exhibit zero winding numbers, indicating 
their topologically trivial nature, as shown by the green line in Fig.~\ref{fig:winding-number-distribution}. 
It is worthwhile to mention that, the winding number confirms the number of edge states. These edge states are 
\textit{non-chiral} in nature, as they are not only restricted to a single edge of the lattice; instead, each 
member of the degenerate pair is localized at the opposite edges of the system. 
\section{Concluding remarks}
\label{sec:conclu}
In conclusion, we have studied the emergence of multiple numbers of nondispersive, completely flat bands in a 
quasi-one-dimensional decorated diamond chain model. It has been shown that, these flat bands can be controlled 
by a proximity effect, which regulates the horizontal diagonal coupling strength in each diamond unit cell. 
The appearance of these flat band states in the system is substantiated by the construction of the compact 
localized states and the computation of the average density of states. It has also been verified that, these 
flat bands are robust against the introduction of small amounts of random disorder in the onsite potential 
distribution of the system. Finally, we have studied the topological features of this lattice model by 
calculating the topologically protected edge states and the winding numbers using the Wilson loop approach. 
These interesting findings for such a simple quantum geometry make it a prospective candidate to be used to 
create a novel, cutting-edge device based on the quantum technologies in the future. Additionally, because 
of the simplicity of the proposed lattice model, it is a promising candidate for experimental 
implementation in a real-life experiment using the laser-induced photonic lattice settings. 
\begin{acknowledgments}
V.T. acknowledges the Govt.\ of India for providing financial support through a NFST research fellowship. 
B.P. acknowledges the Nagaland University for providing funding through a start-up 
research grant for young faculties. B.P. would also like to thank Prof.\ Georges Bouzerar 
and Maxime Thumin from the N\'{e}el Institute for useful discussions on the effect of disorder 
on the flat bands. 
\end{acknowledgments}
\section*{Data Availability Statement}
The data that support the findings of this study are available from the 
corresponding author upon reasonable request. 
\appendix
\section{Effect of vertical diagonal coupling $d_V \neq 0$}
\label{sec:appendix-1}
%
\renewcommand\thefigure{A.\arabic{figure}}    
\setcounter{figure}{0}
\begin{figure}[ht!]
\centering
\includegraphics[clip, width=0.49\columnwidth]{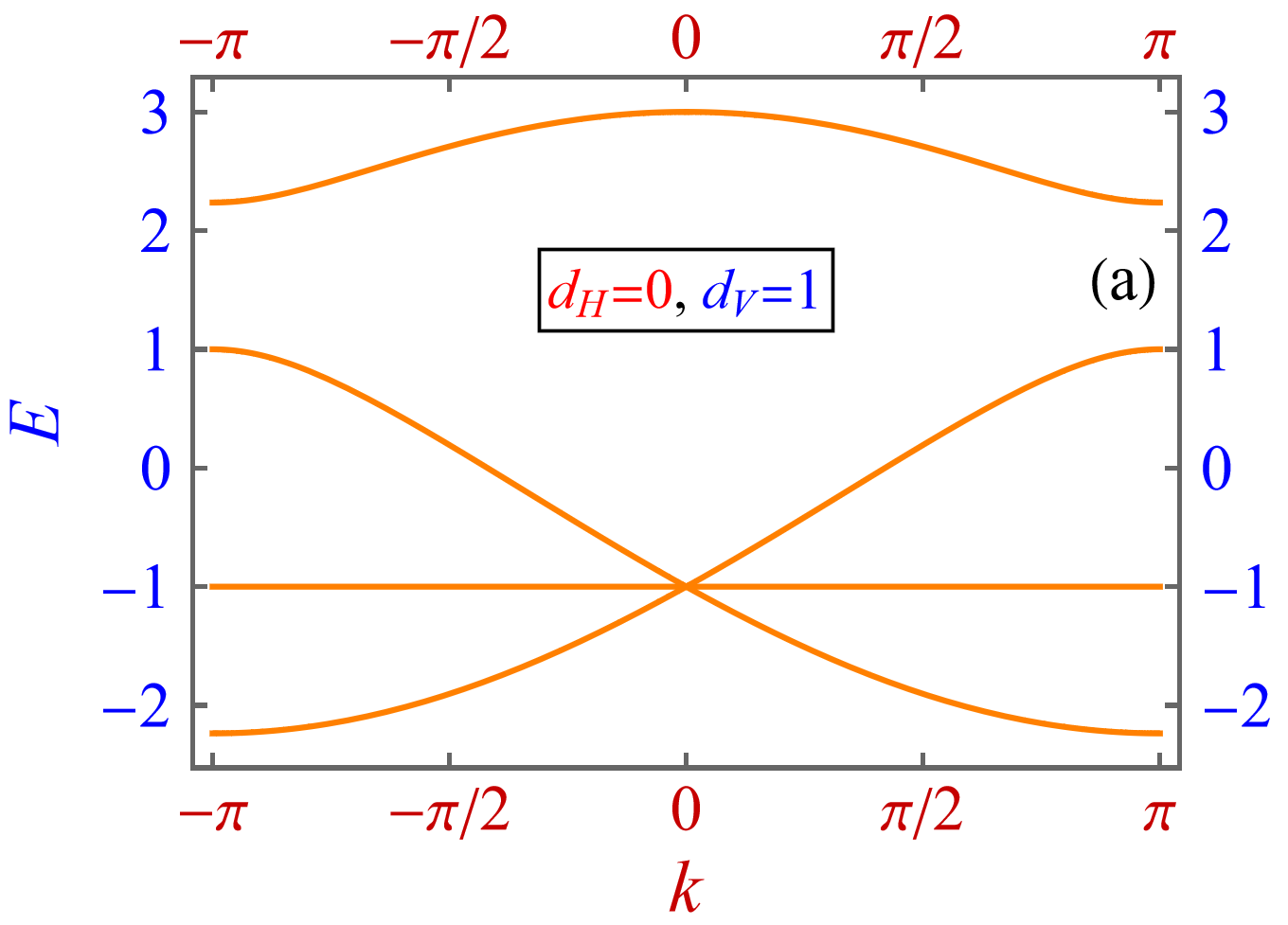}
\includegraphics[clip, width=0.49\columnwidth]{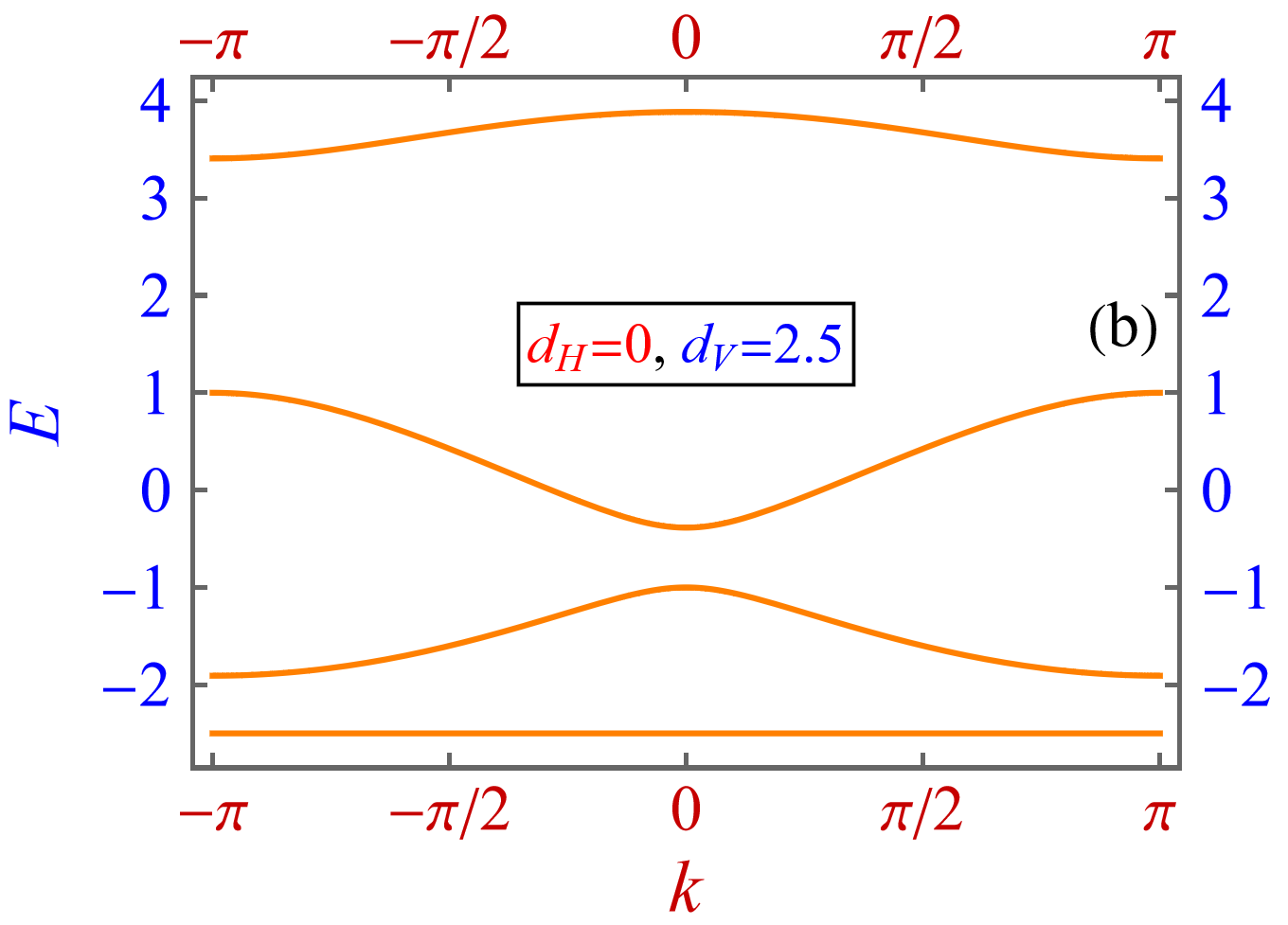}
\caption{Electronic band structure of the decorated diamond lattice model 
with nonzero values of $d_V$. For nonzero finite values of $d_V$, the FB 
appears at $E=-d_V$. (a) For $d_V=1$, a gapless FB appears at $E=-1$ and 
(b) For $d_V=2.5$, a gapped FB appears at $E=-2.5$. The other parameters 
are set to be $d_H=0$ and $\lambda=t=1$.}
\label{fig:Band-structures-with-nonzero-d_V}
\end{figure}
In the main text of the manuscript, we have focused on the effect of the horizontal diagonal coupling $d_H$ 
on the band structure of this lattice model by keeping the vertical diagonal coupling $d_V=0$. However, it 
is a pertinent question to ask what happens for $d_V \neq 0$. This question is addressed in this appendix. We 
find that, for $d_V \neq 0$ (keeping $d_H=0$), a single FB always appears in the band structure at the energy 
$E=-d_V$ (measured in units of $t$). Two sample cases are shown in Fig.~\ref{fig:Band-structures-with-nonzero-d_V}. 
This result is also very interesting, as it gives us the flexibility to set the FB precisely at any desired 
finite value of the energy by tuning the vertical diagonal coupling $d_V$ in the system. 
%
\section{ADOS and IPR with a higher onsite disorder strength $\Delta=0.8$}
\label{sec:appendix-2}
%
\renewcommand\thefigure{B.\arabic{figure}}    
\setcounter{figure}{0}
\begin{figure}[h!]
\centering
\includegraphics[clip, width=0.49\columnwidth]{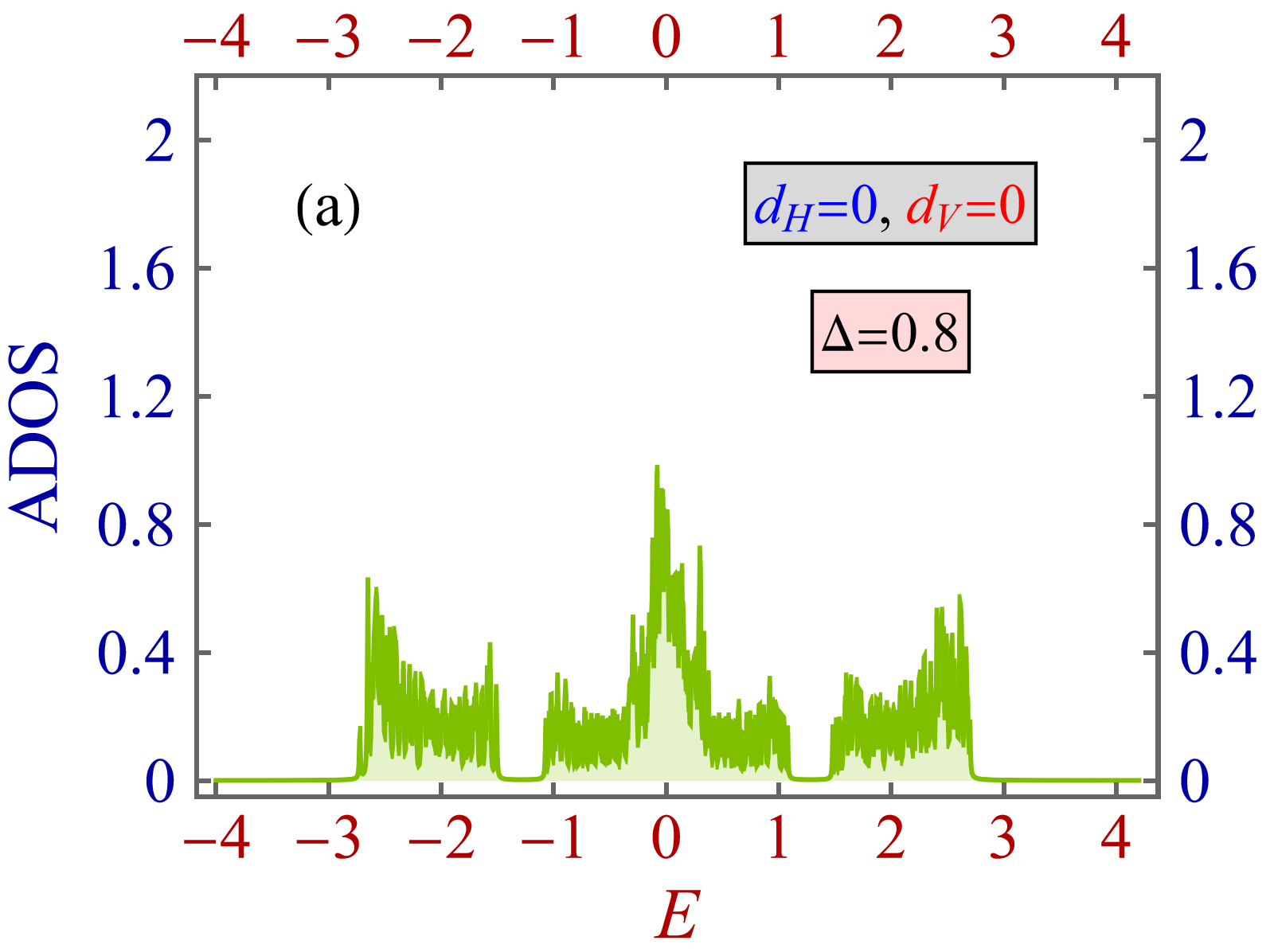}
\includegraphics[clip, width=0.49\columnwidth]{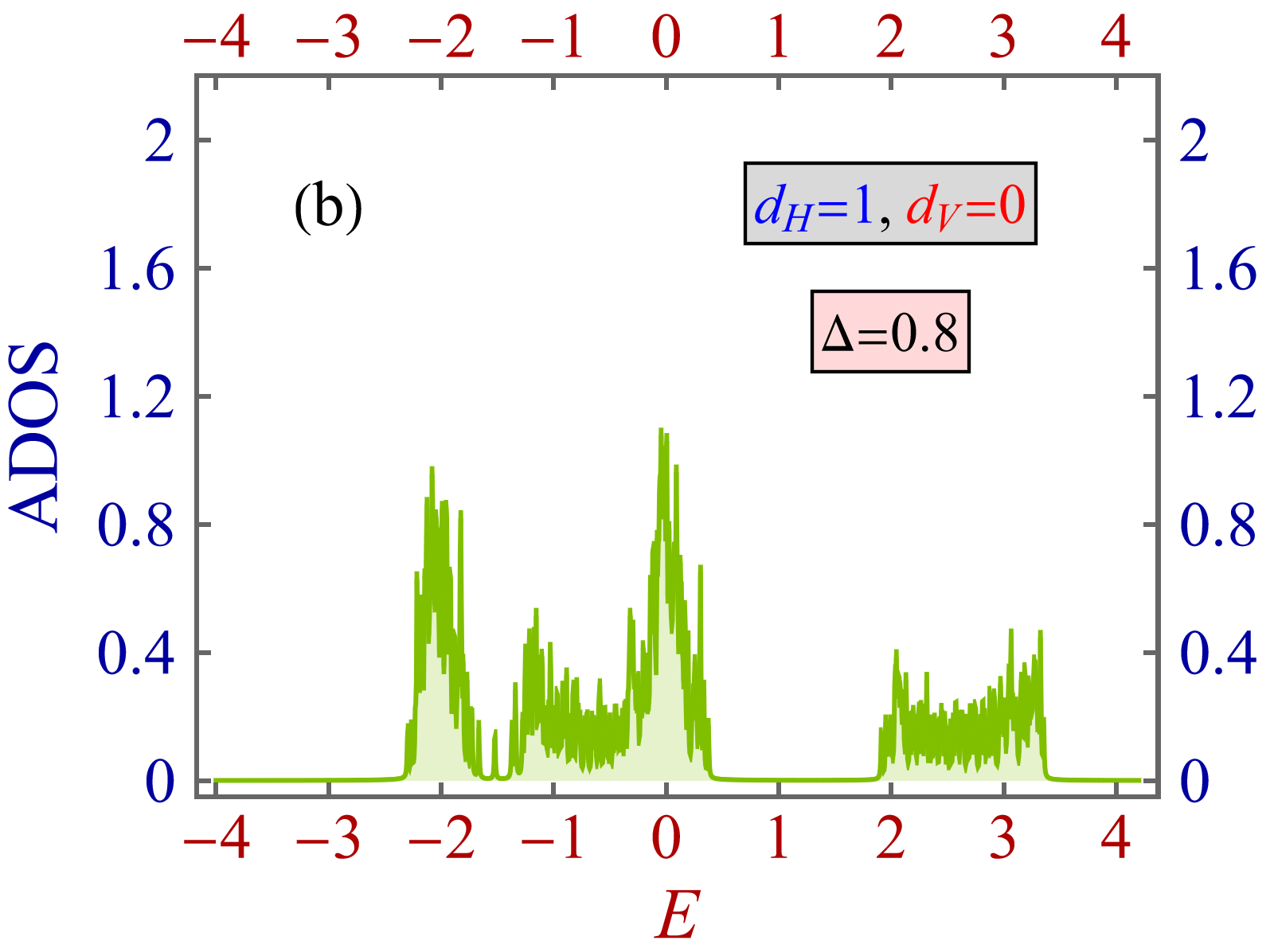}
\vskip 0.4cm
\includegraphics[clip, width=0.49\columnwidth]{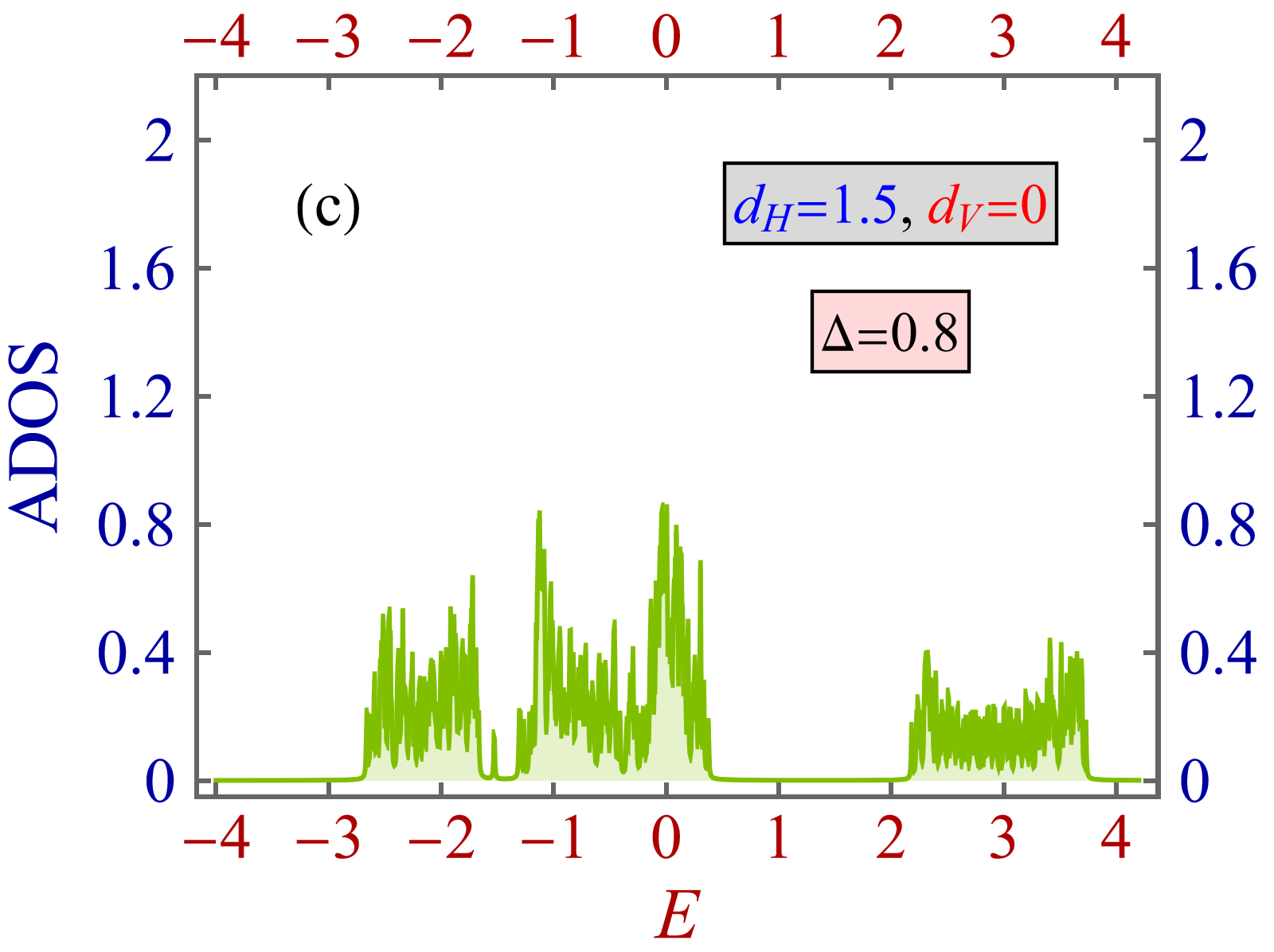}
\includegraphics[clip, width=0.49\columnwidth]{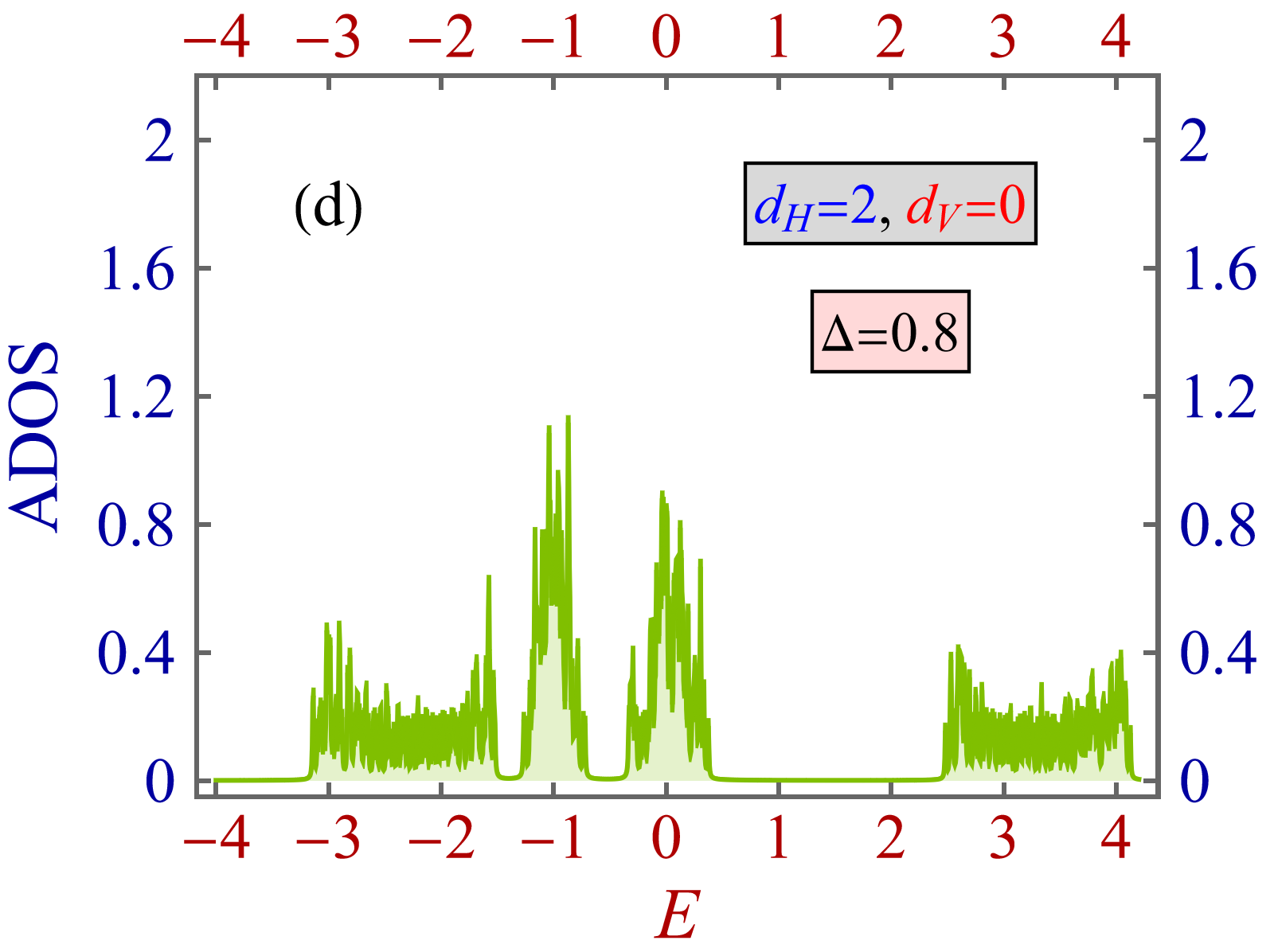}
\caption{Variation of the average density of states (ADOS) as a function of the 
energy ($E$) for the decorated diamond chain for a \emph{large} random onsite 
(diagonal) disorder of strength $\Delta=0.8$ (measured in units of $t$). 
All other parameters are the same as in Fig.~\ref{fig:DOS-with-small-disorder}.}
\label{fig:DOS-with-large-disorder}
\end{figure}
%
As one increases the strength of the onsite disorder $\Delta$ to a higher value, the FB peaks 
start to collapse, and hence we will have a transition from the localized states to the delocalized 
states or moderately localized states in the ADOS spectrum. This is illustrated in Fig.~\ref{fig:DOS-with-large-disorder}, 
where we have chosen a large value of the random onsite disorder strength $\Delta=0.8$ (measured in units of $t$), 
which is $4$-times larger than the value of $\Delta$ taken in Fig.~\ref{fig:DOS-with-small-disorder}.
%
\begin{figure}[h!]
\centering
 \includegraphics[clip, width=0.49\columnwidth]{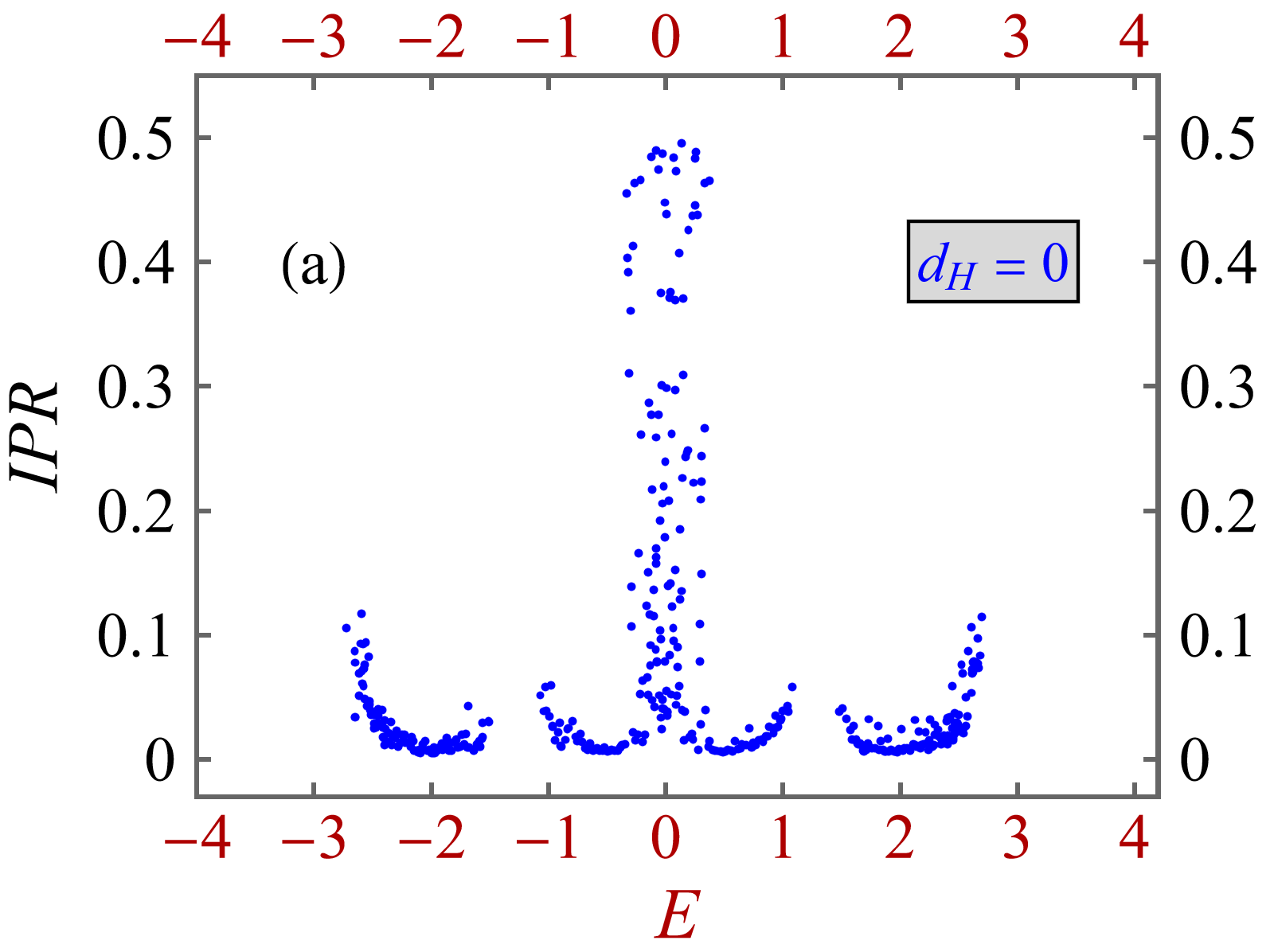}
 \includegraphics[clip, width=0.49\columnwidth]{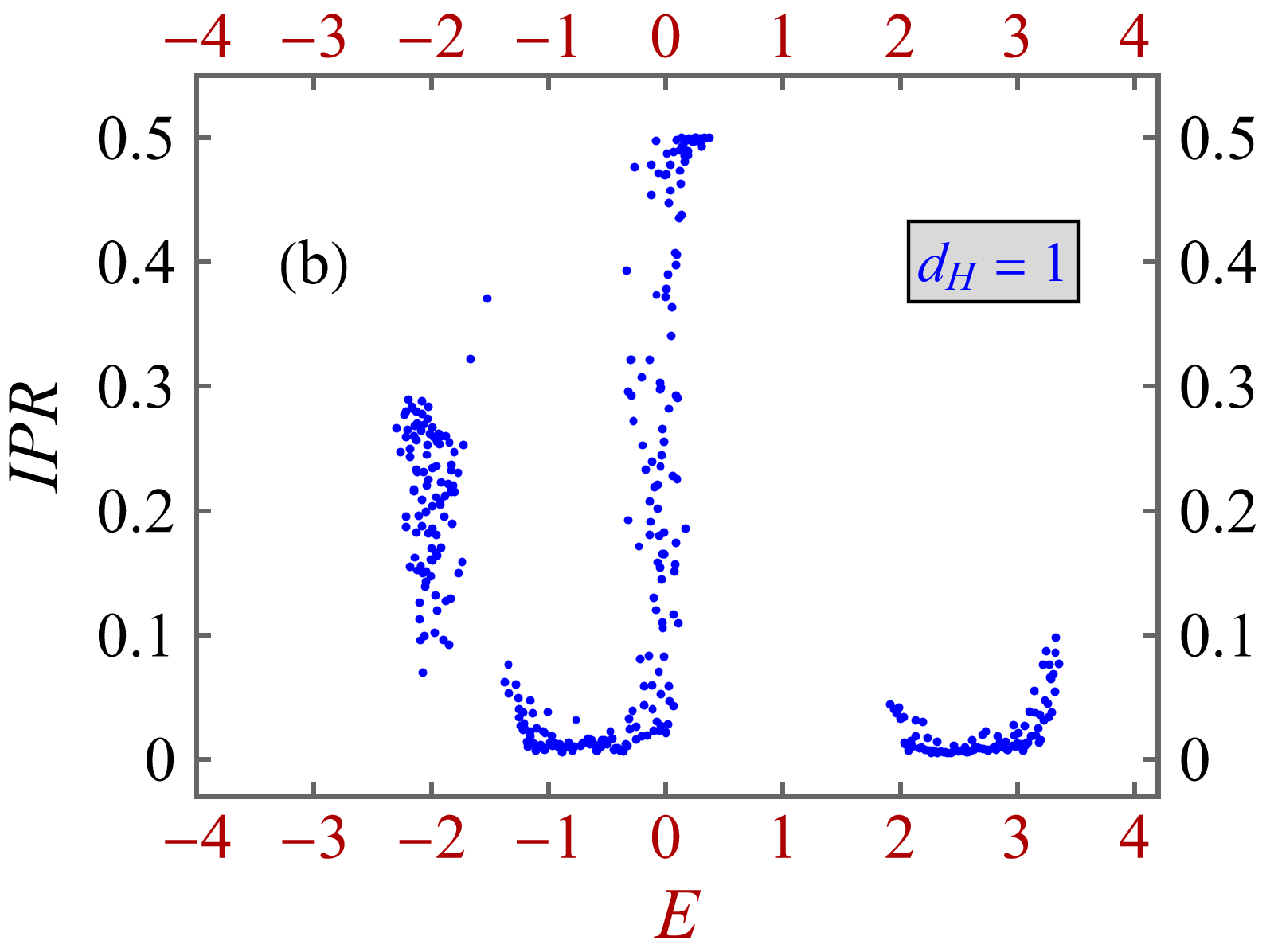}
\vskip 0.4cm
 \includegraphics[clip, width=0.49\columnwidth]{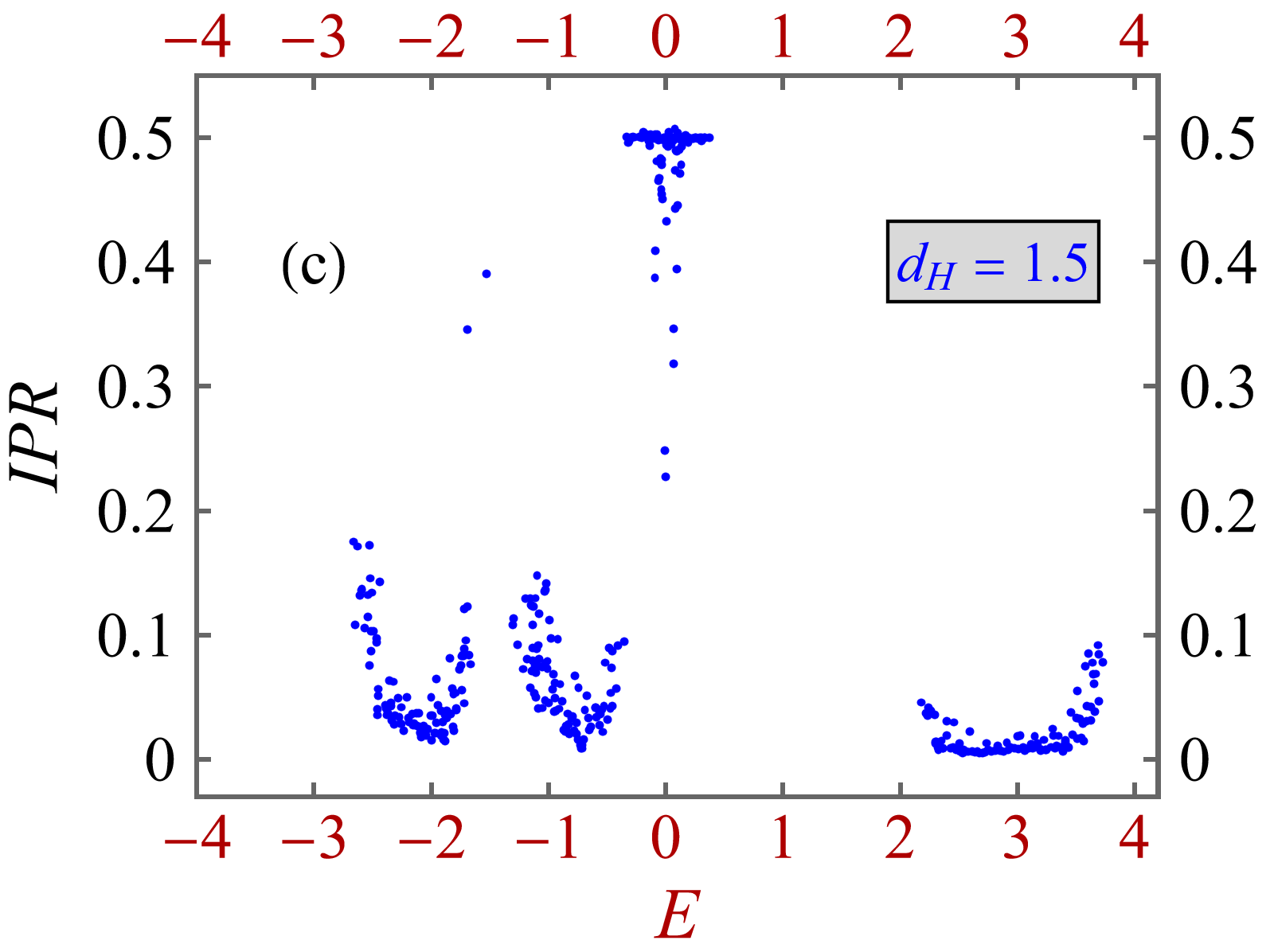}
 \includegraphics[clip, width=0.49\columnwidth]{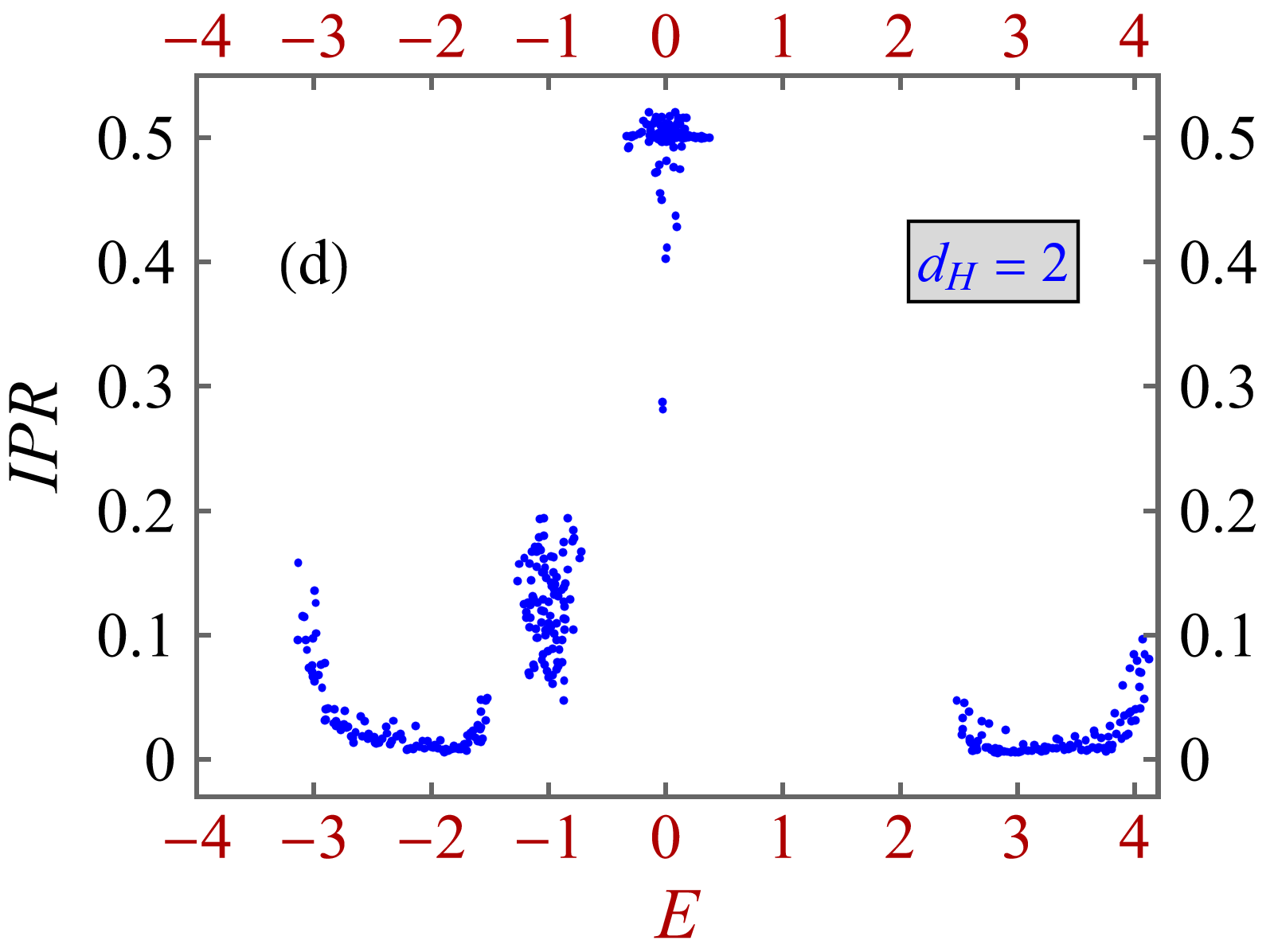}
\caption{Variation of the inverse participation ratio (IPR) as 
a function of the energy ($E$) for the decorated diamond chain of 400 sites with 
a random onsite (diagonal) disorder of strength $\Delta=0.8$ (measured in units of $t$). 
All other parameters are taken to be the same as in Fig.~\ref{fig:DOS-with-large-disorder}.}
\label{fig:IPR-with-large-disorder}
\end{figure}

Now, to further quantify the nature of localization behavior of the wavefunctions, we have 
computed the inverse participation ratio (IPR) in presence of this random onsite disorder. The IPR is 
defined as~\cite{ipr1,ipr2,ipr3}
\begin{equation}
\mathrm{IPR} = \dfrac{\sum\limits_{i} |\psi_i|^4}{\Big(\sum\limits_{i}|\psi_i|^2\Big)^2},
\end{equation}
where the index $i$ runs over all atomic sites in the system. This quantity serves as a measure of the spatial 
extent of the eigenstates — the smaller values correspond to delocalized (extended) states, while the larger values 
indicate the localized states. For a representative disorder strength of $\Delta=0.8$ (measured in units of $t$), 
the calculated IPR is plotted as a function of the energy $E$ in Fig.~\ref{fig:IPR-with-large-disorder}. As evident 
from the figure, the IPR remains close to zero in the extended regime, whereas its value increases up to 
approximately $0.5$ for the moderately localized eigenstates.

\end{document}